\documentclass[referee,useAMS,usenatbib]{mn2e}
\def\lsim{\lower.5ex\hbox{$\; \buildrel < \over \sim \;$}}
\def\gsim{\lower.5ex\hbox{$\; \buildrel > \over \sim \;$}}
\def\be{\begin{equation}}
\def\ee{\end{equation}}

\def\bc{\begin{center}}
\def\ec{\end{center}}
\def\eg{{e.g.}}
\def\etal{{\em et al.}}
\def\ie{{\rm i.e.}}

\def\thetin{\Theta_{\rm in}}
\def\vin{v_{\rm in}}
\def\lin{L_{\rm in}}
\def\rin{r_{\rm in}}
\def\rg{r_{\rm g}}
\def\rci{r_{\rm ci}}
\def\rco{r_{\rm co}}
\def\rsh{r_{\rm sh}}
\def\rmdot{R_{\dot m}}
\def\rout{r_{\rm out}}
\def\hp{$h_p^2$}
\def\jrb{\rm j}
\def\crb{\rm c}
\def\irb{\rm i}
\def\bb{{\rm b}}
\def\etab{{\mbox{$\eta$}}}
\def\chib{{\mbox{$\chi$}}}
\newcommand{\msol}{\mbox{M$_{\odot}$}}
\newcommand{\mbh}{\mbox{$M_{\rm BH}$}}

\usepackage{graphicx}
\usepackage{mathrsfs}
\usepackage{amsmath}
\title[General relativistic viscous accretion disc]
{Estimation of mass outflow rates from viscous relativistic accretion discs around black holes}
\author[Chattopadhyay \& Kumar]
{Indranil Chattopadhyay$^{1}$, Rajiv Kumar$^{1}$\thanks{E-mail: indra@aries.res.in (IC);
rajiv.k@aries.res.in (RK)}\\
$^{1}$Aryabhatta Research Institute of Observational Sciences (ARIES), Manora Peak, Nainital-263002, India\\
}
\begin{document}
\date{}
\maketitle
\label{firstpage}

\begin{abstract}
We investigated flow in Schwarzschild metric, around a non-rotating black hole and obtained self-consistent
accretion - ejection solution in full general relativity. We covered the whole of parameter space in the advective
regime to obtain shocked, as well as, shock-free accretion solution. We computed the jet streamline using von - Zeipel
surfaces and projected the jet equations of motion on to the streamline and solved them simultaneously with
the accretion disc equations of motion. We found that steady shock cannot exist 
{for $\alpha \gsim0.06$} in the general
relativistic prescription, but is lower if mass - loss is considered too.
We showed that for fixed outer boundary, the shock moves closer to the horizon with increasing
viscosity parameter.
The mass outflow rate increases as the shock moves closer to the black hole,
but eventually decreases, maximizing at some intermediate value of shock {location}. The jet terminal speed increases
with stronger shocks, quantitatively speaking, the terminal speed of
jets $v_{{\rm j}\infty} > 0.1$ if $\rsh < 20 \rg$. The maximum of the outflow rate obtained in the general relativistic
regime is less than $6\%$ of the mass accretion rate.
\end{abstract}


\section {Introduction} \label{sec:intro}
 Large amount of radiation emitted by astrophysical objects like microquasars and active galactic nuclei (AGNs) favours the scenario
that such energy output is due to the conversion of gravitational energy of matter into heat and radiation as
it falls into extremely relativistic objects like black holes (BHs). Microquasars are essentially X-ray binaries
and are supposed to harbour a stellar mass BH ($\mbh \sim 10 \msol$ ), while AGNs harbour supermassive BH \ie~
${\mbh} \sim 10^{6-9} \msol$. The radiation emitted by these objects in general contains a relatively low energy
multi-coloured blackbody component
and one or more power-law components in the higher energy limit.
When the accretion disc is in a state, from which the power emitted maximizes
in the higher energy region and the luminosity is low, it is called the low/hard (LH) state.
When the power maximizes in the lower energy level, the disc is luminous and produces multi-coloured blackbody radiation,
it is called the high/soft state (HS). There are many intermediate states (IM) which connects the two.
Along with energetic photons,
AGNs and microquasars also eject highly energetic, collimated and relativistic bipolar jets. Observations
of a large number of microquasars showed that the jets are seen only when the accretion is in the LH
or IM,
but the jet is not seen when the accretion disc is in canonical HS spectral state \citep{gfp03,fbg04,fg14}, \ie  
the jet states are correlated with the spectral states of the accretion disc. 
Such a correlation between
spectral states and jet states cannot be made in AGNs, partly, because of the longer timescale associated with
supermassive BHs and partly, due to possible lack of the periodic repetitions of the outer boundary condition of AGN accretion discs.
However, the fact that timescales in AGNs and microquasars can be scaled by mass \citep{mkkf06}
tells us that the essential physics around super-massive and stellar mass BHs are similar.

The
first popular model of accretion disc around BH was proposed
by \citet{ss73} and \citet{nt73}, and is known as Keplerian disc or standard disc or Sakura-Sunyaev (SS) disc.
It is
characterized by matter rotating with local Keplerian angular velocity, with negligible infall velocity,
and is geometrically thin but optically thick. Being optically thick, each annuli
emits radiation which is thermalized with the matter. Each annulus has different temperature and therefore the spectrum emitted
is a sum of all the blackbody
radiations from each of the annuli, \ie multi-coloured blackbody spectrum. Indeed, the thermal radiation part of
a BH candidate spectrum is well explained by a Keplerian disc. 
Although SS disc was very successful in explaining the thermal component
of the spectrum emitted by BH candidates, but it could not explain the hard powerlaw tail. The inner boundary condition
of the SS disc is quite arbitrary and is chopped off within the marginally stable orbit. The pressure gradient term
and the advection term in SS disc are also poorly treated. It was realized that there should atleast be another component
in the disc, which
would behave like a Comptonizing cloud of hot electrons to produce the hard power-law tail \citep{st80}. 
Moreover, the inner boundary condition of BH dictates that matter crosses its horizon with the speed of light, and that
the angular momentum of the flow close to the horizon needs to be necessarily sub-Keplerian. Therefore, in addition to SS discs,
investigations of accretion in sub-Keplerian regime also gained prominence, such as thick accretion discs
\citep{pw80}, advection-dominated accretion flows or ADAF \citep{nkh97}, advective-transonic regime \citep{lt80,f87,c89}.
All these models start with exactly the same
set of equations of motion i.e., Navier-Stokes equation in strong gravity, but differ in boundary conditions.
For example, if the radial advection term and the pressure gradient term are negligible, azimuthal shear is responsible
for viscosity and the heat dissipated due to viscosity is thermalized locally and efficiently radiated out,
then the resulting disc is the SS disc. On the other hand, if only the advection term is negligible and the cooling
is less efficient, then the model is thick disc. The ADAF and the transonic regime are not subjected
to such confinement, infact, \citet{lgy99} showed that global ADAF is indeed a subset of general
transonic solutions. Recently, by playing around with the viscosity parameter and cooling efficiency in
the computational domain, \citet{gc13} were able to generate both sub-Keplerian advective disc and Keplerian
disc simultaneously. The Keplerian disc gives out soft photons, and sub-Keplerian flow supplies hot electrons,
if the disc has a shock transition. The post-shock disc behaves like a Comptonizing cloud, and produces
the hard power-law photons.

The transonic/advective disc has several advantages. It satisfies the inner boundary condition of the BH, i. e.,
matter crosses the horizon at the speed of light and therefore it is supersonic and sub-Keplerian. It implies that the
existence of
a single sonic point (the position where bulk velocity crosses the local sound speed) is guaranteed around a BH.
However, depending on the angular momentum, there can be multiple sonic points. As a consequence, matter accelerated
through the outer sonic point can be slowed down due to the presence of centrifugal barrier. This slowed down matter
may impede the supersonic matter following it, and may cause shock transition
\citep{f87,c89}. Shock in BH accretion has been found to exist for inviscid flow
\citep{f87,c89,adn15}, dissipative flow \citep{d07,kc13}, and has also been confirmed in simulations \citep{mrc96,lrc11,dcnm14}.
The post-shock region of the disc (PSD), has some special properties. Apart from producing hard powerlaw photons,
it was shown for an inviscid disc
via numerical simulations, that the extra thermal gradient force in the PSD
powers bipolar jets \citep{mlc94,msc96},
and was later
established for
viscous disc as well \citep{lmc98,cd07,dc08,kc13,dcnm14,kcm14}.   
Moreover, since the jet originates from PSD (which extends from few to few tens of Schwarzschild radii)
and not the entire disc, it satisfies the observational
criteria that jets are generated from the inner part of the accretion disc \citep{jbl99,detal12}.

Most of the theoretical studies of accretion on to BHs have been in the domain of pseudo-Newtonian potential (pNp) 
\citep{pw80} and fixed adiabatic index ($\Gamma$) equation of state (EoS) of the flow.
Using pNp gravity potential instead of the Newtonian one has the advantage that, the Keplerian angular momentum
distribution, the location of marginally stable orbit ($r_{\rm m}$), marginally bound orbit
($r_{\rm b}$), or, the photon orbit ($r_{\rm ph}$) can be obtained exactly, as is obtained in general relativity (GR),
but can still remain in the Newtonian regime of physics.
However, according to relativity, matter cannot achieve the speed
of light ($c$) outside the horizon, but, in pNp regime matter velocity exceed $c$ outside the horizon.
The effective potential of a rotating particle is zero on the horizon in GR, however, it is negative infinity
on the horizon if we use pNp. Moreover, in relativity the physics of fluid is different from that of the
particles.
This arises because 
in relativistic equations of motions the thermal term, the angular
momentum term etc{,} couples with the gravity. As a result, for conservative systems, the constants of motion are not the same
in particles and fluids. While in pNp regime, the constants of motion in fluid and particles
are identical. For viscous flow, the shear tensor in relativity is much more complicated and contains many more terms
when compared to the shear tensor in pNp regime. Therefore, solutions of relativistic equations for transonic
accretion discs
around BH have been few \citep[for \eg][]{lt80,l85,f87,c96}
when compared with those in pNp regime and that too in the inviscid limit.
The first consistent viscous advective accretion solution in pure
GR was obtained by \citet{pa97}. They derived the shear tensor from the first principle, and then
approximated it with a simpler but accurate function. For inviscid flow the constants
of motions are the relativistic Bernoulli parameter (${\cal E}=-hu_t$, $h$ is the enthalpy
and $u_t$ is the covariant time component of the four velocity), the accretion rate, angular momentum and the entropy along
a streamline. For viscous flow, except the accretion rate,
none of these are constant along the motion, and constants of motion need
to be determined. The information of the constants of motion
were not used at all by \citet{pa97}, which resulted in a limited class of solutions. 
Moreover, they did not discuss the issue of massloss
either. We would like to rectify that, i. e., to say we would like to obtain all possible accretion solutions
using constants of motion and constants of integration,
as well as, estimate the mass loss from the accretion solution.  

Another limitation of a large body of work on accretion-ejection solutions around
compact objects is that, most of the work has been done assuming a
fixed $\Gamma$ equation of state (EoS), where, $\Gamma$ is the adiabatic index. From classical fluid mechanics, we know that
$\Gamma$ is the ratio of specific heats, 
which turns out to be equal to the constant $5/3$, if
random motions of the constituent particles of the gas are negligible compared to $c$. However, if the
random speeds of the particles is comparable to $c$, then $\Gamma$ is not constant and the EoS
becomes a combination of modified Bessel's function of the inverse of temperature \citep{c39,s57,cg68}. 
It can be trivially shown that the different
forms of the exact EoS obtained by the above three authors are equivalent \citep{vkmc15}.
Moreover, it has been shown that it is unphysical to use fixed $\Gamma$ EoS when the temperature
changes by a few orders of magnitude \citep{t48}. 
The first accretion solution using a relativistic
EoS on to a Schwarzschild BH was obtained by \citet{bm76}. 
\citet{t07} regenerated the solutions of Peitz and Appl,
but also obtained solutions with another form of viscosity using variable $\Gamma$ EoS in Kerr-Schild metric.
However, the EoS used was again for a fluid 
composed of similar particles. Fluids around BH should be fully ionized given the temperature associated
with these fluids, and ionized single species fluid can only be electron-positron flow which cannot exist
for thousands of Schwarzschild radii around the BH.
\citet{bm76} however, hinted how to describe a fluid composed of different particles.
\citet{f87}
in a seminal paper solved accretion solutions in the advective domain for electron-proton flow, and predicted the
possibility of accretion shocks around BH. The inherent problem of using
the exact relativistic EoS in simulation codes is that, it is a ratio of modified Bessels function
which make transformation between primitive variables and state variables non-trivial.
To circumvent this problem we obtained an approximate EoS which is very accurate \citep{rcc06} for single
species fluid, and then extended it to multi-species fluid \citep{c08,cr09,cc11}. 
The adiabatic EoS was also obtained for such a flow by integrating the
entropy generation equation without source terms
\citep{kscc13}. The comparison of Chattopadhyay-Ryu (CR) EoS with an exact one showed negligible
difference between the two \citep{vkmc15}.
The approximate CR EoS was also used in the pNp regime to study dissipative
accretion flow \citep{kc14}, which showed that accretion shocks may exist
for very high viscosity, as well as, high accretion rates. Moreover, depending on these flow parameters
such discs can be of low luminosity, as well as, can emit above the Eddington limit. 
Interesting as it may be, but we know pNp regime can only be considered to be qualitatively correct,
and a general relativistic viscous disc should be considered to fully understand the behaviour of such discs.
Investigations of general relativistic, dissipative, advective accretion discs around BH, described by relativistic EoS
has not been done for multi-species EoS,
in addition, estimation of mass loss from such disc has not been undertaken as well. Apart from the highly non-linear equations
of motion in GR to contend with, it is also a fact that in curved space time, the constant angular momentum surfaces 
are special surfaces called von-Zeipel surfaces \citep[e. g.][and references therein]{c85}.
Jets launched with some angular momentum would follow these surfaces. So an accretion-ejection system in GR is significantly
different from pursuing the same study in pNp regime.
In this paper, we obtain
a simultaneous, self-consistent bipolar jet solution from a general relativistic viscous disc
around a BH, described by multi-species relativistic EoS.

In the next section, we present the equations of motion for the accretion disc and the jet, and also a brief
description of the EoS used. In Section \ref{sec:solproc}, we present the solution procedure of the equations
of motions. In Section \ref{sec:result}, we present the results, and then present our concluding remarks in 
Section \ref{sec:conclusn}.

\section {Assumptions and Equations} \label{sec:equatn}
\begin{figure}
 \centering
 \includegraphics[width=15.0cm]{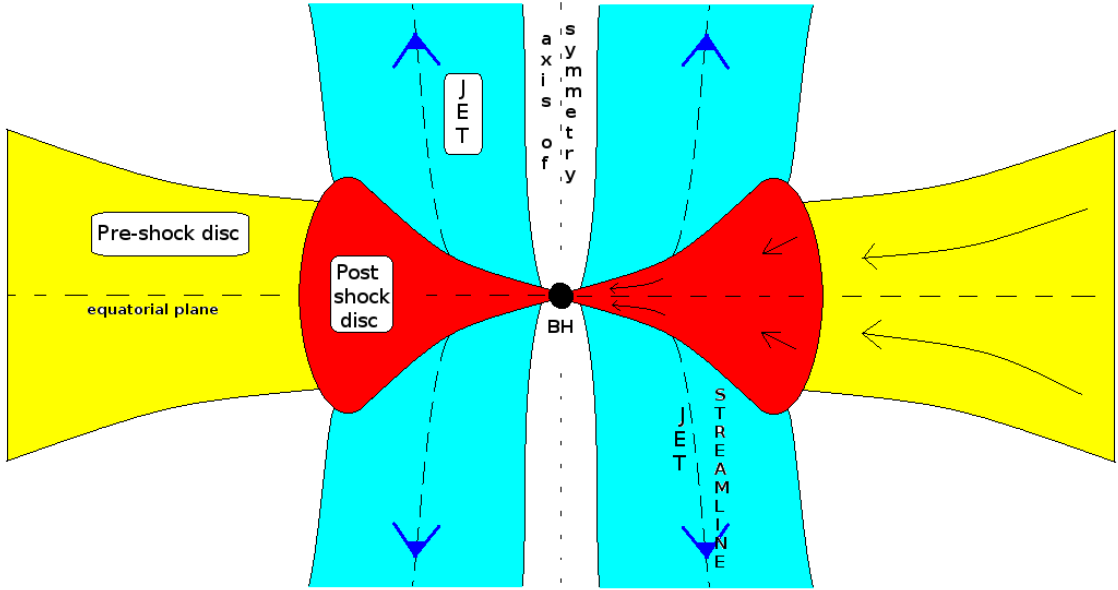}
 \caption{Cartoon diagram of disc-jet system. The arrows show the direction of motion. The disc flow geometry
is on and around the equatorial plane, while the jet flow geometry is about the axis of symmetry. The post-shock disc
{or} PSD and the pre-shock disc are shown. The jet streamline is also mentioned. Here BH stands for the black hole.}
 \label{fig:fig1}
\end{figure}
In this section, we first present the equations of motion
governing the accretion disc and then those governing the matter leaving the
disc as bipolar jets. Although equations of motion for both disc and jets are conservation of four-momentum
and four-mass flux, but since the flow geometry of the disc and that of the jet are different, we will separately present
the two sets of equations. In Fig. \ref{fig:fig1}, a cartoon diagram of the disc jet system is presented.
The accretion disc occupies the region around the equatorial plane, while the jet flows about the axis of symmetry.
The jet geometry is significantly different from the pNp prescription and will be described in Section
\ref{subsec:ouequatn}.

\subsection{Equations governing accretion disc} \label{subsec:inequatn}
The energy momentum tensor for the viscous flow is
\begin{equation}
T^{\mu\nu}=(e+p)u^\mu u^\nu+pg^{\mu\nu}+t^{\mu\nu}, 
\label{emt.eq}
\end{equation}
where $e$, $p$ and $u^\mu$ are the local energy density, local gas pressure and four-velocities, respectively.
The inverse of the metric tensor components is $g^{\mu\nu}$ and Greek indices $\mu, \nu$ represent the space-time coordinates.
Here, $t^{\mu\nu}$ is 
viscous stress tensor and considering it is only the shear that gives rise to the viscosity,
then $t^{\mu\nu}=-2\etab\sigma^{\mu\nu}$, where $\etab$ is
the viscosity coefficient. 
The shear tensor has the general form 
\citep{pa97}
\begin{equation}
\sigma_{\mu\nu}=\frac{1}{2}\left[(u_{\mu;\gamma}h_\nu^\gamma+u_{\nu;\gamma}h_\mu^\gamma)-\frac{2}{3}\Theta_{\rm exp}h_{\mu\nu} \right],
 \label{shear.eq}
\end{equation}
where $h_{\mu\nu}=g_{\mu\nu}+u_\mu u_\nu$ is the projection tensor,  
and $\Theta_{\rm exp}=u_{;\gamma}^\gamma$ is expansion of the fluid world line. Equation (\ref{shear.eq}) can be
rewritten as
\begin{equation}
\sigma_{\mu\nu}=\frac{1}{2}\left[(u_{\mu;\nu}+u_{\nu;\mu}+a_\mu u_\nu+a_\nu u_\mu)-\frac{2}{3}\Theta_{\rm exp}h_{\mu\nu} \right],
 \label{shear1.eq}
\end{equation}
where $a_\mu=u_{\mu;\gamma}u^\gamma$ is the four-acceleration. The covariant derivative of covariant component of
four-velocity is defined as
$u_{\mu;\gamma}=u_{\mu,\gamma}-\Gamma_{\mu\gamma}^\beta u_{\beta}$, 
where $\Gamma_{\mu\gamma}^\beta$ is the
Christoffel symbol. We choose the geometric units where $G=M_{\rm bh}=c=1$ ($G$ is the 
gravitational constant, $M_{\rm bh}$ is the mass of the BH), which has been used in all the
equations, unless mentioned otherwise.

The governing equations of the relativistic fluid are
\begin{equation}
 T_{;\nu}^{\mu\nu}=0, ~~~~~~~~~~~~~~~~~~ (\rho u^\nu)_{;\nu}=0.
 \label{consvfld.eq}
\end{equation}
The relativistic Navier Stokes equation is obtained by projecting the energy momentum conservation along
the $i{\rm th}$ direction i. e.
~$h_{\mu}^iT_{;\nu}^{\mu\nu}=0$ ($i=1, 2, 3$) and can be written as, 
\begin{equation}
[(e+p)u^\nu u_{;\nu}^i+(g^{i\nu}+u^iu^\nu)p_{,\nu}]+h_\mu^it_{;\nu}^{\mu\nu}=0
 \label{NS.eq}
\end{equation}
The energy generation equation or the first law of thermodynamics is $u_\mu T_{;\nu}^{\mu\nu}=0$ and is given by,
\begin{equation}
u^\mu\left[\left(\frac{e+p}{\rho} \right)\rho_{,\mu}-e_{,\mu} \right]=Q^+,
 \label{ege.eq}
\end{equation}
where, $Q^+=t^{\mu\nu}\sigma_{\mu\nu}$ is the viscous heating term and we ignore cooling terms, to stress on the effect of
viscous dissipation. Here $\rho$ is the mass density of the flow and $h$ is the specific enthalpy of the flow,
\begin{equation}
h=\frac{e+p}{\rho}.
 \label{enth.eq}
\end{equation}
We have 
considered only the $r-\phi$ component of relativistic shear tensor.
This would on one hand simplify the equations tremendously, and on the other 
hand would allow us to
directly compare with the plethora of work done with pseudo potentials \citep{bdl08,kc13,kcm14,kc14}. 
The $r-\phi$ component of the shear tensor (equation \ref{shear1.eq}) is written as \citep{pa97}
\begin{equation}
2\sigma_\phi^r=u_{;\phi}^r+g^{rr}u_{\phi;r}+a^ru_\phi+a_\phi u^r-\frac{2}{3}\Theta_{\rm exp}u^ru_\phi.
 \label{rphic.eq}
\end{equation}
Following \citet{pa97}, we neglect derivatives of $u^r, a^r$ and $\Theta_{\rm exp}$ and equation (\ref{rphic.eq})
becomes
\begin{equation}
2\sigma_\phi^r=(g^{rr}+u^ru^r)\frac{du_\phi}{dr}-\frac{2u_\phi}{r}g^{rr}.
 \label{rphi.eq}
\end{equation}
In this paper, we consider only the
simplest BH metric for the accretion disc, namely the Schwarzschild metric, in which the non-zero
metric components are
$$
g_{tt}=-\left(1-\frac{2}{r}\right);~~ g_{rr}=\left(1-\frac{2}{r}\right)^{-1};~~g_{\theta \theta}=r^2;~~
g_{\phi \phi}=r^2{\rm sin}^2\theta.
$$
For accretion, the flow is around the equatorial plane; therefore, the equations are obtained
at $\theta=\pi/2$ and assumed hydrostatic equilibrium along the transverse direction.
With these assumptions, we write down the radial component of Navier Stokes equation (\ref{NS.eq}),
\begin{equation}
u^r\frac{du^r}{dr}+\frac{1}{r^2}-(r-3)u^{\phi}u^{\phi}+(g^{rr}+u^ru^r)\frac{1}{e+p}\frac{dp}{dr}=0,
 \label{rNS.eq}
\end{equation}
the integrated form of the azimuthal component of equation (\ref{NS.eq}),
\begin{equation}
-\rho u^r(L-L_0)=2\etab\sigma_\phi^r,
 \label{phiNS.eq}
\end{equation}
where
 $L=h u_{\phi}=hl$ and $L_0$ are 
the local bulk angular momentum and bulk angular momentum at the horizon of the BH, respectively.
It must be remembered that while $l=u_\phi$ is a conserved quantity in the absence of dissipation for particles,
for fluid $L$ is the corresponding conserved quantity. The specific angular momentum for fluid is therefore
$\lambda=-u_\phi/u_t$, but for particles it is $l$ or $u_\phi$. Moreover,
the radial three velocity is defined as $v^2_{\hat r}=-(u_ru^r)/(u_tu^t)$ and in the local corotating frame
$v^2=\gamma^2_{\phi}v^2_{\hat r}$ \citep{l85}. The associated Lorentz factors being
$\gamma_v=(1-v^2)^{-1/2}$, $\gamma_\phi=(1-v^2_\phi)^{-1/2}$ and the total Lorentz factor is $\gamma=\gamma_v\gamma_\phi$.
Moreover, $v_\phi=\sqrt{-u_\phi u^\phi/u_tu^t}=\sqrt{\Omega \lambda}$, where $\Omega=u^\phi/u^t$.
The hydrostatic equilibrium along the transverse direction gives local disc height expression 
\citep{l94,rh95,pa97},
\begin{equation}
H={\left(\frac{p r^3}{\rho\gamma_{\phi}^2}\right)}^{1/2}.
 \label{hhe.eq}
\end{equation}
The first law of thermodynamics (equation \ref{ege.eq})
\begin{equation}
u^r\left[\left(\frac{e+p}{\rho} \right)\rho_{,r}-e_{,r} \right]=t^{r\phi}\sigma_{r\phi}
\label{ege2.eq}
\end{equation}
Integrating mass-conservation equation, we obtain the expression of the mass accretion rate,
\begin{equation}
 -\dot{M}=4\pi\rho H u^r r.
 \label{mf.eq}
\end{equation}

We can now define the dynamical viscosity coefficient and it is 
 $\etab=\rho \nu$, where the kinematic viscosity is given by $\nu=\alpha a r f_c,
 ~\mbox{$a$ is the sound speed (see equation \ref{pol.eq})}$ $\mbox{and}~f_{c}=(1-v^2)^2$.
Since $\sigma_{r\phi}$ may or may not be equal
to zero on the horizon, with the choice of $f_{c}$ we have made $t^{r\phi}|_{\rm horizon}=0$
\citep[see][for details]{pa97}.
 
The constant of motion can be obtained by integrating equation (\ref{rNS.eq}),
\begin{equation}
 {\rm log}(E)=-\frac{1}{2}{\rm log}(1-v^2)+\frac{1}{2}{\rm log}\left(1-\frac{2}{r}\right)-\int\frac{(r-3)l^2}{r^3(r-2)\gamma_v^2}{\rm d}r
 +\int\frac{1}{e+p}{\rm d}p.
 \label{cmpeq1.eq}
\end{equation}
The last term of equation (\ref{cmpeq1.eq}) with the help of equations (\ref{enth.eq}) and (\ref{ege2.eq}) can be 
written as
\begin{equation}
 \int\frac{1}{e+p}{\rm d}p=\int\frac{1}{h}\frac{{\rm d}p}{\rho}=\int\frac{1}{h}\left[{\rm d}h-\frac{t^{r\phi}\sigma_{r\phi}}{\rho u^r}
{\rm d}r\right].
 \label{cmpeq2.eq}
\end{equation}
Using equation (\ref{phiNS.eq}) and relation $t^{r\phi}=-2\etab\sigma^{r\phi}$ in equation (\ref{cmpeq2.eq}), we get,
\begin{equation}
 \int\frac{1}{e+p}{\rm d}p=\int\frac{1}{h}\left[{\rm d}h+\frac{u^r(L-L_0)^2}{2\nu r(r-2)}{\rm d}r\right].
 \label{cmpeq3.eq}
\end{equation}
Combining equation (\ref{cmpeq3.eq}) in equation (\ref{cmpeq1.eq}) and re-arranging, we get
\begin{equation}
 E=\frac{h\gamma_v\sqrt{1-\frac{2}{r}}}{\exp(X_f)},
 \label{grE.eq}
\end{equation}
where $$X_f=\int\left[\left(\frac{r-3}{r-2}\right)\frac{l^2}{r^3\gamma_v^2}-\frac{u^r(L-L_0)^2}{2\nu hr(r-2)}\right]{\rm d}r.$$
$E$ is the
constant of motion in the presence 
of viscous dissipation and may be called the relativistic Bernoulli constant in the presence of viscosity. It is interesting
to note that in the absence of viscosity, the first term in the parentheses of $X_f$ is ${\rm ln}(\gamma_\phi^{-1})$, and so 
$E({\rm inviscid})=h\gamma_v\gamma_\phi \sqrt{(1-2/r)}=-hu_t={\cal E}$, \ie~ the relativistic Bernoulli constant.
It is indeed intriguing to note that $E$ also has the same dimension of ${\cal E}$, \ie~ of specific energy,
but the former is a constant
of motion while ${\cal E}$ is not. It must be noted that ${\cal E}$ incorporates the information of motion locally, \ie~
motion along
radial and azimuthal direction (quasi-one-dimensional), and the effect of gravity through $-u_t$, while the information of
internal energy is through $h$. Therefore, ${\cal E}$ contains the information of viscous heat dissipation
(it increases where viscosity is effective), but not the angular momentum transport due to viscosity;
as a result, it is not a constant of motion. However, $E$ contains all the information carried by ${\cal E}$,
as well as the information of angular momentum transport, which makes $E$ constant. So it might be {physically more} relevant to
consider $E$ as the specific energy for dissipative flow than ${\cal E}$. Since specific energy expression
in GR is not additive, so all the terms are not apparent; however, a comparison of the constants of motion
for dissipative and inviscid Newtonian flow might be instructive. From \citet{gl04,bdl08} and \citet{kc13,kc14}, one may write down the
grand specific energy or generalized Bernoulli parameter for Newtonian fluid as
$$
E({\rm pNp})=\frac{1}{2}v^2_{\rm pNp}+h_{\rm pNp}-\frac{\lambda^2_{\rm pNp}}{2r^2}
+\frac{\lambda_{\rm pNp}\lambda_{0\rm pNp}}{r^2}-\frac{1}{2(r-1)}.
\eqno{({\rm A})}
$$
The canonical Bernoulli parameter for Newtonian fluid is
$$
{\cal E}({\rm pNp})=\frac{1}{2}v^2_{\rm pNp}+h_{\rm pNp}+\frac{\lambda^2_{\rm pNp}}{2r^2}-\frac{1}{2(r-1)}.
\eqno{({\rm B})}
$$
In the above, the suffix pNp denotes that the flow variables are in pNp regime, $\lambda_{0{\rm pNp}}$
is the specific angular momentum at $\rg$ and the last term on r.h.s of both
the equations (A and B) is the gravity term in pNp. It is clear that while ${\cal E}({\rm pNp})$ contains the
local information of radial {motion} (first {term}), azimuthal motion ($\lambda_{\rm pNp}$),
gravity and the thermal ($h_{\rm pNp}$) terms,
$E$ contains all of them, as well as the angular momentum transport term (third and fourth terms of equation A).
Clearly, if there is no viscosity, then $\lambda_{0\rm pNp}=\lambda_{\rm pNp}$, so $E({\rm pNp}) \rightarrow {\cal E}({\rm pNp})$.
Therefore, one may say $E$ in equation (\ref{grE.eq}) is the constant of motion for viscous, relativistic fluid,
equivalent to the one obtained
in the pseudo-Newtonian limit \citep[e.g.,][]{gl04,kc13,kc14}.

\subsection{Relativistic EoS and the equations of motion:}\label{subsec:reosm}
To solve the equations of motion, we need a closure relation between thermodynamic quantities called the
EoS. In this subsection, we will start by expressing the variables in physical units, and at the end while applying into
equations of motion we will impose the geometric units.
We consider that the fluid is composed of electrons ($e^{-}$),
positrons ($e^{+}$) and protons ($p^{+}$) of varying proportions, but always maintaining the overall charge neutrality: 
$n_{e^-}=n_{p^+}+n_{e^+}$, here $n_s$ is the number density of the $s$th species of the fluid.
The mass density is given by \citet{c08} and \citet{cr09},
\begin{equation}
\rho={\Sigma_{i}}n_{i}m_{i}=n_{e^{-}}m_{e^{-}}\left[2 - {\xi(1 - 1/\chib)}\right]=n_{e^{-}}m_{e^{-}}\tilde{\tau},
\label{rho.eq}
\end{equation}
where, $\chib= m_{e^{-}}/ m_{p^{+}}$, $\xi=n_{p^{+}}/n_{e^{-}}$ is
the composition parameter and $\tilde{\tau}= [2 - {\xi(1 - 1/{\chib})}]$. The electron and proton masses are $m_{e^{-}}$ and $m_{p^{+}}$,
respectively. 
For single temperature flow, the isotropic pressure is given by
\begin{equation}
p = \Sigma_{i}p_{i} = 2n_{e^{-}}kT=2n_{e^-}m_{e^{-}}c^2\Theta=\frac{2\rho c^2\Theta}{\tilde{\tau}}.
\label{p.eq}
\end{equation}
The EoS for multi-species flow is \citep{c08,cr09}
\begin{equation}
e = \Sigma_{i}e_{i} = \Sigma \left[n_{i}m_{i}c^{2} + p_{i}\left(\frac {9p_{i} + 3n_{i}m_{i}c^2}{3p_{i} + 2n_{i}m_{i}c^2}
\right)
\right].
\label{ed.eq}
\end{equation}
The non-dimensional temperature is defined with respect to the electron rest mass energy, $\Theta = kT/(m_{e^{-}}c^2)$.
Using equations (\ref{rho.eq}) and (\ref{p.eq}), the expression of the energy density in equation (\ref{ed.eq}) simplifies to
\begin{equation}
e = n_{e^{-}}m_{e^{-}}c^{2}f = \rho_{e^{-}}c^2f = \frac{\rho f}{\tilde{\tau}},
\label{eos.eq}
\end{equation}
where
$$
f = (2-\xi) \left[1 + \Theta \left(\frac {9\Theta + 3}{3\Theta + 2}\right)\right] + \xi \left[\frac{1}{\chib} + \Theta
\left(\frac {9\Theta + 3/\chib}{3\Theta + 2/\chib}\right)\right].
$$
The expressions of the polytropic index, the adiabatic index and the sound speed are given as,
\begin{equation}
N = \frac{1}{2} \frac{df}{d\Theta};~ \Gamma = 1 + \frac {1}{N},~{\rm and}~ a^2=\frac{\Gamma p}{e+p}=\frac{2\Gamma \Theta}
{f+2\Theta}.
\label{pol.eq}
\end{equation}
Integration of first law of thermodynamics (equation \ref{ege2.eq}) by assuming adiabatic flow ($Q^+=0$) and using the EoS
(equation \ref{eos.eq}),
gives us the adiabatic relation of multi-species relativistic flow \citep{ck13,kscc13},
\begin{equation}
\rho={\cal{K}} ~\mbox{exp}(k_3)~ \Theta^{3/2}(3\Theta+2)^{k_1}(3\Theta+2/\chib)^{k_2}, 
\label{peos.eq}
\end{equation}
where $k_1=3(2-\xi)/4, k_2=3\xi/4$ and $k_3=(f-\tilde{\tau})/(2\Theta)$
and ${\cal{K}}$ is the constant of entropy. Equation (\ref{peos.eq}) is the generalized version of $p={\cal K}\rho^\Gamma$.
Combining equations 
(\ref{peos.eq}) and (\ref{mf.eq}), we get the expression of entropy accretion rate,
\begin{equation}
{\dot {\mathcal{M}}}=\frac{\dot{M}}{4\pi{\cal K}}=\mbox{exp}(k_3) \Theta^{3/2}(3\Theta+2)^{k_1}
(3\Theta+2/\chib)^{k_2}Hru^r.
\label{ent.eq}
\end{equation}

Re-arranging equations (\ref{rNS.eq}-\ref{mf.eq}) with the help of equations (\ref{rphi.eq}), (\ref{enth.eq}), (\ref{rho.eq}),
(\ref{p.eq}) and (\ref{eos.eq}) in geometric units, we present the spatial derivative of flow variables
$v, l$ and $\Theta$, 
\begin{equation}
\frac{dv}{dr}= \frac{\cal{N}}{\cal{D}},
\label{dvre.eq}
\end{equation}
where 
\begin{eqnarray*}
 &&{\cal{N}}=-\frac{1}{r(r-2)}+(\frac{r-3}{r-2})\frac{l^2}{r^3\gamma_v^2}+\frac{2a^2}{\Gamma+1} \\
&& \times \left[\frac{\widetilde{\tau}u^r(L-L_{0})^2}{8\nu r(r-2)(N+1) \Theta}+ 
 \frac{5r-8}{2r(r-2)}-\frac{l^2}{r^2\gamma^2}\left(\frac{1}{l}\frac{dl}{dr}-\frac{1}{r}\right) 
 \right] \\
&&{\cal D}= \gamma_{v}^{2}\left[v-\frac{2a^{2}}{\Gamma+1}\left(\frac{l^2}{r^2\gamma^2} v+ \frac{1}{v}\right)\right].
\end{eqnarray*}
Here, ${\cal D}$ contains an extra term $l^2v/(r^2\gamma^2)$  compared to the inviscid case \citep{cc11}. There
is $\gamma_\phi$ term in the expression of disc height (equation \ref{hhe.eq}). The radial derivative of 
equation (\ref{mf.eq}) implies that the radial derivative of the specific angular momentum will be non-zero, which causes
the extra term to appear. There are many height prescriptions \citep{l94,rh95,pa97}, and choice of any one of them
apart from the one used, will not affect the result qualitatively.
Then, 
 \begin{equation}
\frac{dl}{dr}=\left[- \frac{u^{r}(L-L_{0})}{\nu (1- \frac{2}{r})}+ \frac{2l}{r} \right] (1-v^2).
\label{dlre.eq}
\end{equation}
Moreover,
\begin{eqnarray}\nonumber
\frac{d\Theta}{dr} =- \frac{\widetilde{\tau}u^r(L-L_{0})^2}{2\nu r(r-2)(2N+1)}
- \frac{2 \Theta}{2N+1} ~~~~~~~~~~~~~~~~~ \\ \times \left[\frac{5r-8}{2r(r-2)}+
\gamma_v^2 \left(\frac{1}{v}+v\frac{l^2}{r^2\gamma^2} \right) 
\frac{dv}{dr} - \frac{l^2}{r^2 \gamma^2} \left \{\frac{1}{l} \frac{dl}{dr}- \frac{1}{r} \right \} \right].
\label{dthre.eq}
\end{eqnarray}
These differential equations are integrated by using fourth order Runge Kutta numerical method with the help of 
using critical point conditions and l$^{\prime}$Hospital rule at critical point. 

\subsubsection{Sonic point equations} \label{subsubsec:soniceqn}
Mathematical form of critical point equation is ${\rm d}v/{\rm d}r={\cal N}/{\cal D}=0/0$, which gives two equations as,
\begin{equation}
\left[1-\frac{2}{\Gamma_{\crb}+1}\left(\frac{a_{\crb}^2l_{\crb}^2}{r_{\crb}^2\gamma_{\crb}^2} + \frac{a_{\crb}^2}{v_{\crb}^2}\right)
\right]=0
 \label{de.eq}
\end{equation}
and
\begin{eqnarray}
-\frac{1}{r_{\crb}(r_{\crb}-2)}+\left(\frac{r_{\crb}-3}{r_{\crb}-2}\right)\frac{l_{\crb}^2}{r_{\crb}^3\gamma_{v_{\crb}}^2}+
 \frac{2a_{\crb}^2}{\Gamma_{\crb}+1}~~~~~~~~~~~~~~~~~~~~~~~~~~~~~~~~
 \label{nu.eq}
 \\ \nonumber \times \left[\frac{\widetilde{\tau}u_{\crb}^r(L_{\crb}-L_{0})^2}{8\nu_{\crb}r_{\crb}(r_{\crb}-2)N_{\crb}\Gamma_{\crb} 
 \Theta_{\crb}}+ 
 \frac{5r_{\crb}-8}{2r_{\crb}(r_{\crb}-2)}+\frac{l_{\crb}^2}{r_{\crb}\gamma_{\crb}^2}\left(\frac{u_{\crb}^r(L_{\crb}-L_0)}{\nu_{\crb}
 l_{\crb}\gamma_{v_{\crb}}^2(r_{\crb}-2)}-
 \frac{1-2v_{\crb}^2}{r_{\crb}^2}\right) \right]=0.
\end{eqnarray}
Here, the subscript `${\crb}$' denotes 
the same physical quantities described in equations (\ref{dvre.eq}-\ref{dthre.eq}), but evaluated at the location of
the critical point. The velocity gradient on the sonic point, \ie~   $({\rm d}v/{\rm d}r)_{\crb}$, is obtained by
employing l$^{\prime}$Hospital rule. 

\subsubsection{Relativistic shocks for viscous flow} \label{subsubsec:sok}
The relativistic shock conditions were first obtained by \citet{t48}, which
for viscous flow in the presence of mass-loss are
\begin{equation}
\dot{M}_+=\dot{M}_--\dot{M}_o
 \label{mfc.eq}
\end{equation}
\begin{equation}
 [\Sigma h\gamma_v^2 v v+W]=0
 \label{momfc.eq}
\end{equation}
\begin{equation}
 [\dot{J}]=0
 \label{anmomf.eq}
\end{equation}
\begin{equation}
 [\dot{E}]=0
 \label{ef.eq}
\end{equation}
where, $\dot{J}=\dot{M}L_0=\dot{M}(L-2\nu\sigma_{\phi}^r/u^r),~\dot{E}=\dot{M}E,~ \Sigma=2\rho H~\mbox{and}~{\rm W}=2pH$. 
We have solved four shock conditions (\ref{mfc.eq}-\ref{ef.eq}) simultaneously, where viscous shear tensor
($\sigma_\phi^r$) is 
continuous across the shock and we obtained the relation between pre-shock (suffix `$-$')
and post-shock (suffix `$+$') flow variables,
\begin{equation}
 L_-=L_++(2\sigma_\phi^r|_+)\left[\frac{\nu_+}{u_+}-\frac{\nu_-}{u_-}\right]; 
 ~h_-^\prime u_-^2-{\rm k}_1u_-+2\Theta_-=0;~ {\rm k}_2-\exp({X_f}_-)h_-^\prime{\gamma_v}_-=0,
\label{sok.eq}
\end{equation}
where, ${\rm k}_1=(1-R_{\dot{m}})(h_+^\prime u_+^2+2\Theta_+)/u_+,~R_{\dot{m}}={\dot M}_o/{\dot M}_-,
~{\rm k}_2=\exp({X_f}_+)h_+^\prime{\gamma_v}_+,~h^\prime=(f+2\Theta)~
\mbox{and}~~u=v\gamma_v$. Here, ${X_f}_-=(f_l/f_{\gamma})^2{X_l}_++f_uf_L^2{X_{L}}_+/(f_{\nu}f_h),~
{X_l}_+=\int(\frac{r-3}{r-2})\frac{l_+^2}{r^3{\gamma_v}_+^2}{\rm d}r, ~{X_L}_+= -\int\frac{u_+^r(L_+-L_0)^2}{2\nu_+ h_+r(r-2)}{\rm d}r$,
~~~$f_l=l_-/l_+,~f_{\gamma}={\gamma_v}_-/{\gamma_v}_+,~f_u=u_-^r/u_+^r,~f_L=(L_--L_0)/(L_+-L_0),~f_{\nu}=\nu_-/\nu_+,~f_h=h_-/h_+,
~\mbox{and}~~{X_f}_+={X_l}_++{X_L}_+$. From equation (\ref{phiNS.eq}), viscous shear tensor can be written as
$2\sigma_\phi^r|_+=-u_+(L_+-L_0)/\nu_+$. 

\subsection{Outflow equations} \label{subsec:ouequatn}
The jet being tenuous, we idealize it to be inviscid; therefore, the energy momentum
tensor of jet fluid should be ideal. The general form of the equations of motion would be similar (equation \ref{consvfld.eq});
however, the geometry is entirely different (see Fig. \ref{fig:fig1}).
For the jet we define,
\begin{equation}
 \vartheta^i=\frac{u^i_{\jrb}}{u^t_{\jrb}} ~~~\mbox{and}~~~ \vartheta_i=-\frac{u_{i{\jrb}}}{u_{t{\jrb}}},
 \label{ol.eq}
\end{equation}
where $i=(r, \theta, \phi)$ and `{\jrb}' implies jet quantities and should not be confused with vector or tensor components.
Here, $\vartheta^i$ and $\vartheta_i$ are the component of `transport' velocity (also called
as coordinate velocity) and the respective momentum per unit inertial mass \citep{c85}. The azimuthal
three-velocity of the jet is defined as $v_{\phi~{\jrb}}=(\vartheta_\phi \vartheta^\phi)^{1/2}=(\Omega_{\jrb}
\lambda_{\jrb})^{1/2}$, where $\lambda_{\jrb}$,
the specific angular of the jet, is constant along the flow. The three-velocity of the jet along the stream line
is given by
$v^2_p=\vartheta_r\vartheta^r+\vartheta_\theta \vartheta^\theta$. The surfaces of constant angular momentum for jets
in GR are VZS where the von Zeipel parameter is constant \citep{kja78,c85}. 
The von Zeipel parameter is defined as
\begin{equation}
Z_\phi={\left(\frac{\vartheta_\phi}{\vartheta^\phi}\right)^{1/2}}=\left(-\frac{g^{tt}}{g^{\phi \phi}} \right)^{1/2}
=\frac{r_{\jrb}~{\rm sin}\theta_{\jrb}}{(1-2/r_{\jrb})^{1/2}}.
 \label{VZp.eq}
\end{equation}
Equation (\ref{VZp.eq}) defines the streamline.
The angular momentum of jets would be related to the von Zeipel parameter \citep{c85}
 \begin{equation}
  \vartheta_\phi=c_\phi Z_\phi^n,
  \label{vonzr.eq}
 \end{equation}
where $c_\phi$ and $n$ are some constant parameters. Using equation (\ref{vonzr.eq})
along with EoS (equation \ref{eos.eq}), the definitions of $h$ (equation \ref{enth.eq}) and $Z_\phi$
(equation \ref{VZp.eq})
while  integrating the jet equations of motion gives us the constant of motion of the jet,
which is similar to the Bernoulli parameter along the streamline of the jet,
\begin{equation}
 {\Re}_{\jrb}=-h_{\jrb}u_{t{\jrb}}[1-c_\phi^2Z_\phi^{(2n-2)}]^\beta,
 \label{jtconst.eq}
\end{equation}
where $u_{t{\jrb}}=-(1-2/r_{\jrb})^{1/2}\gamma_{\jrb},~\gamma_{\jrb}=\gamma_{v{\jrb}}
\gamma_{\phi {\jrb}},~\gamma_{v{\jrb}}=1/\sqrt{(1-v_{\jrb}^2)},~
\gamma_{\phi {\jrb}}=1/\sqrt{(1-c_\phi^2Z_\phi^{(2n-2)})} $, $v_{\jrb}=\gamma_{\phi {\jrb}} v_p$
and $\beta=n/(2n-2)$. 
The mass outflow equation can be written as,
\begin{equation}
\dot{M}_o=\rho_{\jrb} u^p_{\jrb} {\cal{A}}_{\jrb},
 \label{mof.eq}
\end{equation}
where $\rho_{\jrb}, u^p_{\jrb}=\sqrt{g^{pp}}\gamma_{v{\jrb}}v_{\jrb}$ and ${\cal A}_{\jrb}$ are jet mass 
density, jet four-velocity along the VZS 
and area of jet cross-section, respectively. The expression of $g^{pp}=1/h^2_p$ is defined in Appendix \ref{app:gpp}.
And similar to the accretion disc equations, we can also derive the entropy-outflow rate for the jet,
and is defined as
\begin{equation}
\dot{\cal M}_{\jrb}=\frac{\dot{M}_o}{2\pi {\cal K}}= \mbox{exp}(k_3)~ \Theta^{3/2}_{\jrb}(3\Theta_{\jrb}+2)^{k_1}
(3\Theta_{\jrb}+2/\chib)^{k_2} u^p_{\jrb} \frac{{\cal{A}}_{\jrb}}{2 \pi}.
 \label{entroj.eq}
\end{equation}
If there are no shocks in jets, then $\dot{\cal M}_{\rm j}$ will remain constant along the streamline.
The differential form of equation (\ref{jtconst.eq}) with the help of equations (\ref{mof.eq}) and (\ref{peos.eq}) and after
some manipulations is obtained as
 \begin{equation}
  \frac{dv_{\jrb}}{dr_{\jrb}}=\frac{\frac{a_{\jrb}^2}{{\cal A}_{\jrb}}\frac{d{\cal A}_{\jrb}}{dr_{\jrb}}
  -\frac{a_{\jrb}^2}{h_p}\frac{dh_p}{dr_{\jrb}}-\frac{1}{r_{\jrb}(r_{\jrb}-2)}}
  {v_{\jrb}\gamma_{v{\jrb}}^2[1-\frac{a_{\jrb}^2}{v_{\jrb}^2}]}=\frac{{\cal N}_{\jrb}}{{\cal D}_{\jrb}}
  \label{dvj.eq}
 \end{equation}
and
\begin{equation}
 \frac{d\Theta_{\jrb}}{dr_{\jrb}}=-\frac{\Theta_{\jrb}}{N_{\jrb}}\left[\frac{\gamma_{v{\jrb}}^2}
 {v_{\jrb}}\frac{dv_{\jrb}}{dr_{\jrb}}+\frac{1}{{\cal A}_{\jrb}}\frac{d{\cal A}_{\jrb}}{dr_{\jrb}}
 -\frac{1}{h_p}\frac{dh_p}{dr_{\jrb}}\right].
 \label{dthj.eq}
\end{equation}
Here, expression of ${\cal A}_{\jrb}$ is defined in equation (\ref{jtar.eq}) in Section \ref{subsec:jtcrimout}.
It is to be noted that
$({\rm d}{\cal A}_{\jrb})/({\cal A}_{\jrb}{\rm d}r_{\jrb})=(r_{\jrb}-1)/[r_{\jrb}(r_{\jrb}-2)]$ and
$({\rm d}h_p)/(h_p{\rm d}r_{\jrb})
=({\rm d}h_1)/(h_1{\rm d}r_{\jrb})-({\rm d}h_2)/(h_2{\rm d}r_{\jrb})-{\rm tan}\theta_{\jrb}({{\rm d}\theta_{\jrb}}/{{\rm d}r_{\jrb}})-1/[r_{\jrb}(r_{\jrb}-2)]$.
Here,
$h_1=1+{{\rm tan}^2\theta_{\jrb}(r_{\jrb}-3)^2}/[{r_{\jrb}(r_{\jrb}-2)}], h_2=h_3^2+h_4^2{\rm tan}^4\theta_{\jrb}(r_{\jrb}-3)^2/(r_{\jrb}-2)^2$, 
~~${\rm d}h_1/{\rm d}r_{\jrb}=-\theta_{\jrb}^\prime {\rm tan}\theta_{\jrb}[(6-r_{\jrb})/r_{\jrb}+(r_{\jrb}-3)\theta_{\jrb}^\prime {\rm tan}\theta_{\jrb}], ~~
{\rm d}h_2/{\rm d}r_{\jrb}=h_3(2-{\rm sin}2\theta_{\jrb}\theta_{\jrb}^\prime)+
h_4(r_{\jrb}-3){\rm tan}^4\theta_{\jrb}[(r_{\jrb}-3)\{1+({\rm sin}2\theta_{\jrb}+4h_4/{\rm sin}2\theta_{\jrb})\theta_{\jrb}^\prime\}+
h_4/(r_{\jrb}-2)]/(r_{\jrb}-2)^2, ~~h_3=(2r_{\jrb}-2-{\rm sin}^2\theta_{\jrb}), ~~h_4=(r_{\jrb}-4+{\rm sin}^2\theta_{\jrb})$
and from differentiation of eq. (\ref{VZp.eq}), we get ${\rm d}\theta_{\jrb}/{\rm d}r_{\jrb}=\theta_{\jrb}^\prime=
-{\rm tan}\theta_{\jrb}(r_{\jrb}-3)/[r_{\jrb}(r_{\jrb}-2)]$.
\subsubsection{Jet sonic point}\label{subsubsec:jtsonicpt}
From the definitions, jet critical point conditions are obtained from equations (\ref{dvj.eq}) and (\ref{dthj.eq}) as,
\begin{equation}
 {\cal N}_{\jrb}=0~~~\Rightarrow ~~~a_{\jrb\crb}^2=\frac{1/[r_{\jrb\crb}(r_{\jrb\crb}-2)]}{[\frac{1}{{\cal A}_{\jrb\crb}}
 \frac{{\rm d}{\cal A}_{\jrb\crb}}{{\rm d}r_{\jrb\crb}}-
 \frac{1}{h_p}\frac{{\rm d}h_p}{{\rm d}r_{\jrb}}]},
 \label{njc.eq}
\end{equation}
and
\begin{equation}
 {\cal D}_{\jrb}=0~~~\Rightarrow ~~~M_{\jrb\crb}^2=\frac{v_{\jrb\crb}}{a_{\jrb\crb}},
 \label{djc.eq}
\end{equation}
where, subscript `{\crb}' denotes flow values at critical point. And the velocity gradient at the critical point
is obtained by l$^{\prime}$Hospital's rule.

\section{Solution procedure} \label{sec:solproc}
We first solve for the accretion solution and once the accretion solution is obtained, we iteratively find the jet
solution from the accretion solution.
Since, close to the horizon, gravity dominates all other physical processes, so the infall time-scale of matter
will be smaller than viscous time-scale or any other time-scales. In other words, very close to the
horizon{,} matter is almost falling freely and $E\simeq{\cal E}$. It may be remembered from Section \ref{subsec:inequatn}
that $E$ is the generalized relativistic Bernoulli parameter
in the presence of viscosity and ${\cal E}$ is the canonical
relativistic Bernoulli parameter. In steady state, for inviscid flow ${\cal E}$ is a constant of motion and
for viscous flow $E$
is a constant of motion. Therefore,
at a distance $r_{\rm in}\rightarrow r_g$, $v_{\rm in}
=\delta\sqrt{2/r_{\rm in}}$. Here, $r_g=2r_s=2GM_{B}/c^2$, $r_{\rm in}=2.001r_s$ and
$\delta<1$. We start by assigning $\delta=1$ in $\vin$, and obtain $\thetin$ and $L_0$. With these values, we integrate
equations (\ref{dvre.eq}), (\ref{dlre.eq}) and (\ref{dthre.eq}) outwards. If the
ensuing solution does not satisfy critical point conditions (equations \ref{de.eq}
and \ref{nu.eq}), we reduce $\delta$ and repeat the procedure till the accretion critical points are obtained
and thereby fixing the value of $\delta$. 

\subsection{Method to find $L_0$}\label{subsec:calL0}
We have provided four flow parameters ($E, \xi, \alpha$ and $\lambda_{\rm in}$ or $\lin$) and by using $v_{\rm in}$, we can 
calculate $\thetin$ from relativistic Bernoulli equation ${\cal E}=-hu_t$. Since we know
$u_t~[=-\sqrt{(1-2/r)}~\gamma]$  from 
$v_{\rm in}$, $\lambda_{\rm in}$ and $E={\cal E}$ at $r=r_{\rm in}=2.001r_s$, so enthalpy ($h$) can be expressed as cubic equation in
$\Theta$ from 
enthalpy equation (\ref{enth.eq}), which is
\begin{equation}
 X_3\Theta^3+X_2\Theta^2+X_1\Theta+X_0=0,
 \label{cubth.eq}
\end{equation}
where, $X_3=72\chib, ~X_2=3[16(\chib+1)-3\chib\tilde{\tau}X_c],
~X_1=2[10-3\tilde{\tau}(X_c-1)(\chib+1)],~ X_0=-4\tilde{\tau}(X_c-1)$ and 
$X_c=-{\cal E}/u_t$. Equation (\ref{cubth.eq}) gives three real roots but two are negative and only one is positive,
so we used 
positive root and is symbolized as $\thetin$. Now, $L_0$  can be calculated from equation 
(\ref{grE.eq}) by assuming $E={\cal E}$ at $r_{\rm in}$. 
Since we assume $E={\cal E}=-hu_t$ close to the horizon, therefore, from equation (\ref{grE.eq})
at $r=\rin$ we have $\gamma_{\phi}\exp(X_f)=1$. This condition is written as,
\begin{equation}
 -\frac{1}{\gamma_{\phi}}\frac{d\gamma_{\phi}}{dr}=\left[\left(\frac{r-3}{r-2}\right)\frac{l^2}{r^3\gamma_v^2}-\frac{u^r(L-L_0)^2}
 {2\nu hr(r-2)}\right].
 \label{cond.eq}
\end{equation}
Simplifying the above equation with the help of equations (\ref{dvre.eq}) and (\ref{dlre.eq}), we get a
quadratic equation in $L_0$, given by
\begin{equation}
 b_2L_0^2+b_1L_0+b_0=0,
 \label{am0.eq}
\end{equation}
where, $b_2=u^r[{\tilde{\tau}vv_{\phi}^2\wp}/(4\Theta DN\Gamma)+{1}/{h}]/[2\nu r(r-2)]$, $b_1=-2\lin b_2-a_1$ and $b_0=b_2\lin^2
+a_1\lin+a_0$. 
Here, $a_1=[u^rv_{\phi}^2(v-\wp/v)]/[\nu\gamma_v^2l(1-2/r)D]$, $a_0=vv_{\phi}^2[-1+(5r-8)\wp/2+(r-3)\gamma_{\phi}^2\wp(v_{\phi}^2+1/v^2)+
(r-2)(1-\wp/v^2)(1-2/\gamma_{\phi}^2)]/[r(r-2)D]$, 
$\wp=2a^2/(\Gamma+1)$ and $D=[v-\wp(v_{\phi}^2 v+ {1}/{v})]$. Equation (\ref{am0.eq}) gives two real roots, one is greater
than $\lin$ and 
other less than $\lin$. Since viscosity transports angular momentum outward, so second root,
which is less than 
$\lin$, is the correct solution.

To summarize, we have obtained asymptotic values of $\vin~{\rm and}~\thetin$ at $r_{\rm in}=2.001$ and $L_0$ or $\lambda_0$ 
on the horizon by using three flow parameters, $E, \alpha$, $\lin$ and $\xi$ to fix the EoS, so that
we can integrate equations 
(\ref{dvre.eq} - \ref{dthre.eq}) simultaneously outwards from $\rin$.
It is to be noted that only correct values of $\vin,~ \thetin ~{\rm and}~ L_0$ will produce a transonic solution.

\subsection{To find critical point and shock locations in disc} \label{subsec:crpt}
Initially, a tentative accretion solution is obtained without considering mass-loss from the disc. 
We obtain the transonic solution iteratively, \ie,  to say, for a given set of $(E, \alpha$, $\lin)$, there exists
a unique set of $\vin,~\thetin~{\rm and}~ L_0$ which will pass through a certain critical point ($r_{\rm c}$).
Once we obtain $r_{\crb}$, we integrate outwards to obtain global solution. Gravity induces one sonic point
or critical point. Rotation induces multiple sonic points. If the first sonic point obtained is close
to the horizon, we call it inner sonic point $\rci$. If the transonic solution is monotonic, then
there are no other sonic points. Once we get one sonic point, we continue to search for other sonic points.
Up to three sonic points can be obtained, in which the inner ($\rci$) and the outer ($\rco$) sonic points
are X-type and are physical sonic points as flow actually passes through these sonic points.
The middle sonic point is unphysical because flow actually does not pass through it, since the $({\rm d}v/{\rm d}r)_{\crb}$
at middle
sonic point is complex. For viscous fluid, the middle sonic point is spiral type.

For flows going through $\rci$, we check for the shock conditions equations (\ref{sok.eq}), initially assuming
${\dot M}_{\rm o}=0$, and compute the pre-shock flow variables (\ie ~$v_-,~a_-,L_-$). We integrate with $v_-,~a_-,L_-$
along the supersonic branch and check whether solution passes through the outer sonic point or $\rco$.
The location of the jump $\rsh$, for which the supersonic branch starting with $v_-,~a_-,L_-$
goes through $\rco$ is the shock location. When there is a shock, then the entropy of the flow through $\rco$
is less than the entropy of the flow through $\rci$, \ie $~{\dot {\cal M}_{\rm o}}< {\dot {\cal M}_{\irb}}$.

\subsection{To find jet critical point and mass outflow rate}\label{subsec:jtcrimout}

While $E~(/{\cal E})$ is the constant of motion along equatorial plane for viscous (/inviscid) accretion solution,
however, away from the equatorial plane, the constant of motion is given by equation (\ref{jtconst.eq}) which is constant
along the jet stream-line defined by equation (\ref{VZp.eq}). Numerical simulations show that the post-shock disc
is the jet base \citep{mrc96,dcnm14}. Numerical simulations also
show that the angular momentum at the top
of the PSD (the base of the jet) is about 20-30\% less than from the equatorial plane, so without losing generality
we consider at the base $\lambda_{\jrb}=2\lambda/3$, and the location of the jet base $x_\bb=(\rci+\rsh)/2$.
We estimate $\Re$ at $x_\bb$ on the disc surface and the jet is launched with the same modified Bernoulli
parameter, \ie~ $\Re_{\jrb}=\Re (x_\bb)$.
The modified Bernoulli parameter ($\Re_{\jrb}$) depends 
on constants $n$ and $c_\phi$ apart from its local flow variables.
Interestingly, the entropy of the jet also depends on these two parameters. Keeping same $\Re_{\jrb}$,
but by changing $n$ and $c_\phi$, iteratively, we obtain the ${\dot {\cal M}}_{\jrb}$
which admits the transonic jet solutions,
with the help of equations (\ref{njc.eq}) and (\ref{djc.eq}) for particular values of $n>0$. 
Since only a fraction of matter escapes as jets, so 
$\dot{\cal M}_{\jrb}$ should be less than local disc entropy at $x_b$ but greater than the disc pre-shock entropy.
Following the above constraint, $c_\phi$ and $n$ would be related by
$c_\phi=Z_\phi^n/\lambda_{\jrb}$.

Once we know the jet solution it is easy to define the relative mass outflow rate,
\begin{equation}
R_{\dot{m}}=\frac{\dot{M}_0}{\dot{M}_-}=\frac{1}{[\dot{M}_+/\dot{M}_0+1]}.
 \label{rmd1.eq}
\end{equation}
The jet base cross-sectional area, perpendicular to tangent of the stream line at $r_{\jrb}$ is,
\begin{equation}
{\cal A}_{\jrb}={\cal A}_\bb \left(\frac{r_{\jrb}}{r_\bb}\right)^2 {\rm sin}\theta_{\jrb},
 \label{jtar.eq}
\end{equation}
where, ${\cal A}_\bb={\cal A}_\bb^\prime {\rm sin}\theta_\bb$ and
${\cal A}_\bb^\prime=2\pi(r_{\bb 0}^2-r_{\bb{\irb}}^2)$ are area along the 
accretion cylindrical radial coordinate and area along the spherical radial coordinate, respectively. Here,
$r_\bb=\sqrt{x_\bb^2+h_\bb^2}, \theta_\bb={\rm sin}^{-1}(x_\bb/r_\bb), r_{\bb{\irb}}=x_{\bb{\irb}}/sin\theta_\bb, r_{\bb 0}=
x_{\bb 0}/{\rm sin}\theta_\bb, 
x_\bb=(\rci+\rsh)/2, x_{\bb{\irb}}=\rci$ and $x_{\bb 0}=\rsh$. Here, $\theta_{\jrb}={\rm sin}^{-1}
(Z_\phi\sqrt{1-2/r_{\jrb}}/r_{\jrb})$ and 
$Z_\phi=r_\bb {\rm sin}\theta_\bb/\sqrt{(1-2/r_\bb)}$.
Now the equation (\ref{rmd1.eq}) with the help of equations (\ref{jtar.eq}), (\ref{mof.eq}) and (\ref{mf.eq}) can be written as,
\begin{eqnarray}
 R_{\dot{m}}=\frac{1}{\left[{(4\pi H_+r_+\rho_+u_+^r)}/{({\cal A}_{{\jrb}\bb}\rho_{{\jrb}\bb}u_{{\jrb}\bb}^p)}+1 \right]} \\ \nonumber
    = \frac{1}{\left[\varSigma{(R_{\rm A}R\varXi)}^{-1}+1 \right]},
 \label{rmd.eq}
\end{eqnarray}
where $\rho_{{\jrb}\bb}=\rho_\bb {\rm exp}(-7x_\bb/(3h_\bb))/h_\bb^2, u_{{\jrb}\bb}^p=\sqrt{g^{pp}}\gamma_{v\bb}v_{{\jrb}\bb}$ and 
${\cal A}_{{\jrb}\bb}={\cal A}_\bb {\rm sin}\theta_\bb$ are jet base density, four-velocity at jet base and jet base area, respectively.
Moreover, $R_{\rm A}={\cal A}_{{\jrb}\bb}/(4\pi H_+r_+)$, $R=(u_-^r)/(u_+^r)$ the compression ratio, $\varSigma=\rho_+/\rho_-$
o, the density jump across the accretion shock 
and $\varXi=(\rho_{{\jrb}\bb}u_{{\jrb}\bb}^p)/(\rho_-u_-^r)$ or the ratio of the relativistic mass flux of the pre-shock accretion flow
and the jet base, respectively. It is to be noted that $\varXi$ measures the upward thrust imparted by the shock
through the compression ratio.

Once the jet solution is obtained for a particular accretion shock solution, we compute the relative mass outflow rate
or $ R_{\dot m}$, and feed it back to the shock conditions (equation \ref{sok.eq}) and retrace the steps
mentioned in Sections \ref{subsec:calL0} and \ref{subsec:crpt} to find a new $\rsh$. Then
from this new $\rsh$ we find a new jet solution and new $R_{\dot m}$ (Section \ref{subsec:jtcrimout}).
We continue these
iterations till the shock location converges and then we obtain a self-consistent accretion-ejection
solution around BHs in full general relativistic regime.

\section{Results} \label{sec:result}

In this paper, we obtained jet solution from accretion solutions. In other words, we supplied accretion
disc parameters $E, \alpha$, $\lin$
and $\xi$ to fix the EoS of the relativistic flow, obtained accretion and jet solutions simultaneously.
However, in the following subsection
we will first present all possible accretion solution and then in the next subsection we will present
the accretion-ejection solutions. The location of the outer boundary of the accretion disc is $10^5\rg$ for
totally sub-Keplerian disc and/or wherever the angular momentum distribution
achieves the local Keplerian value.

\subsection{Inflow solutions} \label{subsec:insoln}

\begin{figure}
 \centering
 \includegraphics[width=9.0cm]{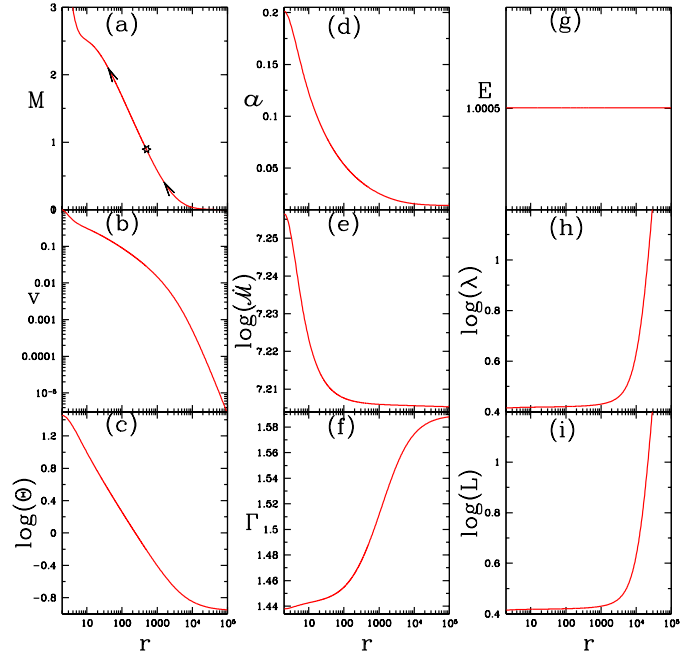}
 \caption{Variation of accretion Mach number $M$ (a), bulk velocity $v$ (b), dimensionless temperature 
 $\Theta$ (c), sound speed $a$ (d), entropy accretion rate $\dot{\cal M}$ (e), accretion adiabatic index $\Gamma$
(f), generalized relativistic Bernoulli parameter $E$ (g), specific angular momentum $\lambda$ (h) and bulk angular momentum
$L$ (i). The sonic point is indicated by the star mark in panel (a).
The accretion disc parameters are $E=1.0005,~L_0=2.6,~\alpha=0.01$ and $\xi=1.0$.}
 \label{fig:fig2}
\end{figure}
In Fig.(\ref{fig:fig2}), we plot the accretion solution for $E=1.0005,~L_0=2.6,~\alpha=0.01$. We choose $\xi=1.0$, until
specified otherwise. Various flow variables plotted are the Mach number $M=v/a$ (a), $v$ (b), $\Theta$ (c), $a$ (d),
$\dot{\cal M}$ (e), $\Gamma$ (f), $E$ (g), $\lambda$ (h) and $L$ (i). The disc parameters were such that it produces
a single outer-type sonic point. While $\Gamma$ varies from semi-relativistic to relativistic values ($1.437 <\Gamma < 1.59$),
the constant of motion $E$ is indeed a constant. The entropy also increases due to viscous dissipation. And
the angular momentum is transported outwards.

\begin{figure}
 \centering
 \includegraphics[width=9.0cm]{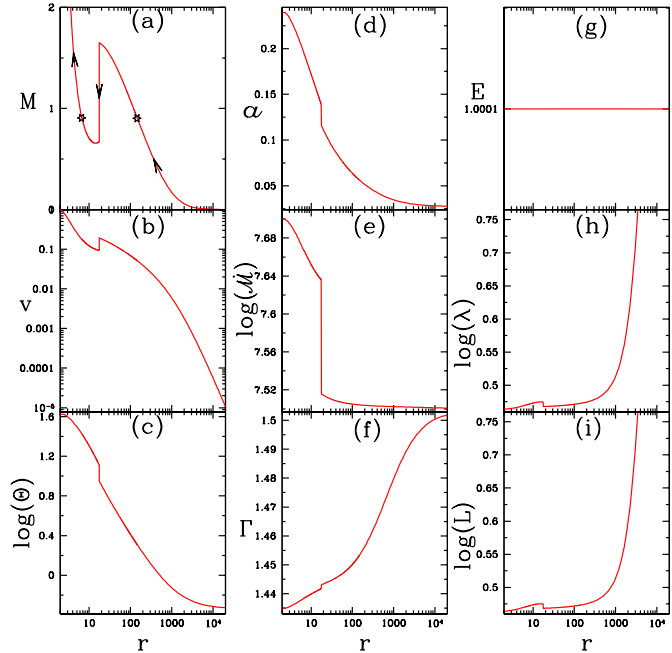}
 \caption{Variation of accretion Mach number $M$ in plot (a), bulk velocity $v$ in plot (b), dimensionless temperature 
 $\Theta$ in plot (c), local sound speed
 $a$ in plot (d), entropy accretion rate $\dot{\cal M}$ in 
 plot (e), accretion adiabatic index $\Gamma$ in plot (f), general relativistic Bernoulli parameter $E$ in plot (g), 
 specific angular momentum 
 $\lambda$ in plot (h) and bulk angular momentum $L$ in plot (i) are shown in this figure. Here, vertical jump shows the
 location 
 of shock, which is $r_s=51.19$ and the two star marks in panel (a) indicate the X-type sonic points.
The accretion 
 disc parameters are $E=1.0001, L_0=2.91, \alpha=0.01$ and $\xi=1.0$.}
 \label{fig:fig3}
\end{figure}

In Fig.(\ref{fig:fig3}),
we have shown typical {shocked} accretion solution and variation of various flow quantities with radial 
distance, for a different value of $E~(=1.0001)$ and $L_0~(=2.91)$ while keeping the viscosity and the nature of
the fluid similar to the previous figure. Since $E$ is a constant of motion in the viscous relativistic disc, and $L_0$
is a constant of integration, so changing these two disc parameters 
{is} equivalent to changing the inner
boundary condition of the accreting flow. It is to be noted that, {the solution in 
Fig. (\ref{fig:fig2})} is similar to a Bondi type solution
\citep[\ie~ low angular momentum flow through an outer critical point $\rco$; see][]{b52}.
So accretion flow is not decidedly monotonic or shocked,
it depends on the boundary condition of the flow.

\begin{figure}
 \centering
 \includegraphics[width=10.0cm]{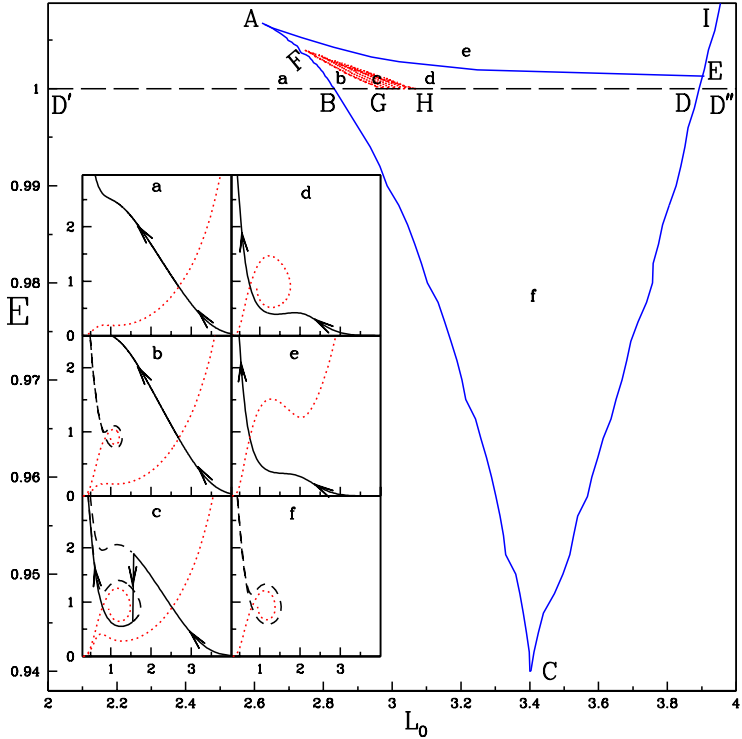}
 \caption{Division of parameter space ($E, L_0$) on the basis of number of critical points and corresponding solutions
 topologies
 [Mach number, $M$, versus radial distance, ${\rm log}(r)$ plots in panels a, b, c, d, e and f]. In this figure
 viscosity 
 parameter, $\alpha=0.01$ and composition parameter, $\xi=1.0$.}
 \label{fig:fig4}
\end{figure}

In Fig. (\ref{fig:fig4}), we obtain a parameter space
of $E$ and $L_0$ for $\alpha=0.01$ and $\xi=1$, and demarcate the regions which will give transonic solutions
with single sonic points, multiple sonic points and shocked solutions. For all $E,~L_0$ values in the domain
ABD$^{\prime}$, angular momentum is low and all {possible} solutions {in this domain}
will possess a single outer-type sonic point
similar to Bondi flow (typical Mach number variation: panel a). The region BGFB is with a bit more angular momentum
and the inner sonic point ($\rci$) appears, although the accreting matter still flows through $\rco$
into the BH (typical solution: panel b). Since the entropy of $\rci$ is higher for these values of $E~{\rm and}~L_0$,
so oscillating shock is a distinct possibility. Solutions in the domain GFHG admit steady-state shock in accretion
solutions and thereby joining the solutions through outer and inner sonic points (typical solution:
panel c). In the domain HFADEH, the angular momentum is much higher, multiple sonic points still
exist, but the accreting matter prefers to flow into the BH through $\rci$ because ${\dot {\cal M}}_{\rm i}>
{\dot {\cal M}}_{\rm o}$
(typical solution: panel d). For solutions from the region AEI, the angular momentum is so
large that matter falls with very low inflow velocity, and becomes transonic only close
to the horizon, and therefore possess an inner-type sonic point  only (typical solution:
panel e). Solutions from the domain BDCB are bound through out and do not produce global transonic solutions
(typical solution: panel f).
{The solid curves within panels (a) --- (f) indicate 
physical solutions, which accreting matter actually follows}.
The dashed part of the solution indicates those which are viable solutions but matter
do not choose. The dotted curves in the panels
show also transonic solutions which have wind-type boundary conditions (low $v$
close to horizon and high $v$ at large distances). However, these so-called wind-type solutions
should not be confused with 
proper wind or outflow solutions, since these solutions are defined only on the equatorial plane.

\begin{figure}
 \centering
 \includegraphics[width=9.0cm]{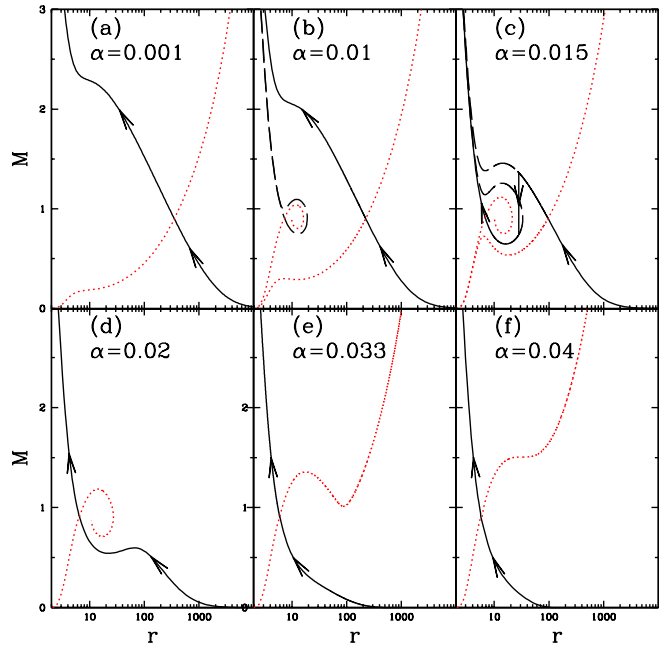}
 \caption{Variation of $M$ with $r$ for different viscosity parameters marked in each panel.
For all panels, $E=1.001,~L_0=2.85,~{\rm and}~\xi=1.0$.}
 \label{fig:fig5}
\end{figure}

All possible accretion solutions can also be produced even if the viscosity is varied
for a given value of $E,~L_0$ and $\xi$.
In Fig. (\ref{fig:fig5}a), we obtain a Bondi-type solution for a low-$L_0$ and low-viscosity
($\alpha=0.001$) solution. We know viscosity transports angular momentum outwards, but low $\alpha$
means the angular momentum remains low at the outer edge too. Such low angular momentum does not produce a strong centrifugal
barrier and therefore produces a shock-free Bondi-type solution with a single, outer-type sonic point.
Keeping the same inner boundary condition, we increase the viscosity to $\alpha=0.01$
and multiple sonic points appear in Fig. (\ref{fig:fig5}b). Higher viscosity for the same values of $L_0$
implies higher angular momentum at larger distances. Gravity ensures a single sonic point;
however, for higher angular momentum flow, the effect of gravity is impeded by rotation at distances
of few tens of $\rg$, while gravity dominating at distances further away, and also very close to the horizon.
This causes multiple sonic points to form. Increasing to $\alpha=0.015$ and keeping
the same inner boundary
condition, steady accretion shock is obtained in Fig. (\ref{fig:fig5}c). Higher $\alpha$ also ensures
even higher $\lambda$ of the disc, thus enhancing the centrifugal barrier. This causes the supersonic matter to be slowed
down and eventually forms a shock. For even higher viscosity
$\alpha=0.02$, the solution through the inner sonic point opens up as shown in Fig. (\ref{fig:fig5}d).
Increasing the viscosity
even further, monotonic accretion solution is obtained (Figs. \ref{fig:fig5}e, f).
If the angular momentum increases beyond a certain limit, then the accreting matter becomes rotation dominated,
and becomes supersonic only very close to the horizon. Therefore, accreting matter does not pass through outer sonic point
(if present), and falls on to the BH through the inner sonic point. 
{Hence}, there exist
two critical $\alpha$ for such boundary conditions, where the lower value
of it would initiate the shock and the higher one will remove it. 
{Such dependence of the nature of accretion solution on viscosity parameter
have been studied in the pNp regime before \citep{c96,cd07,kc13,kc14},
but not in the GR regime.}

\begin{figure}
 \centering
 \includegraphics[width=9.0cm]{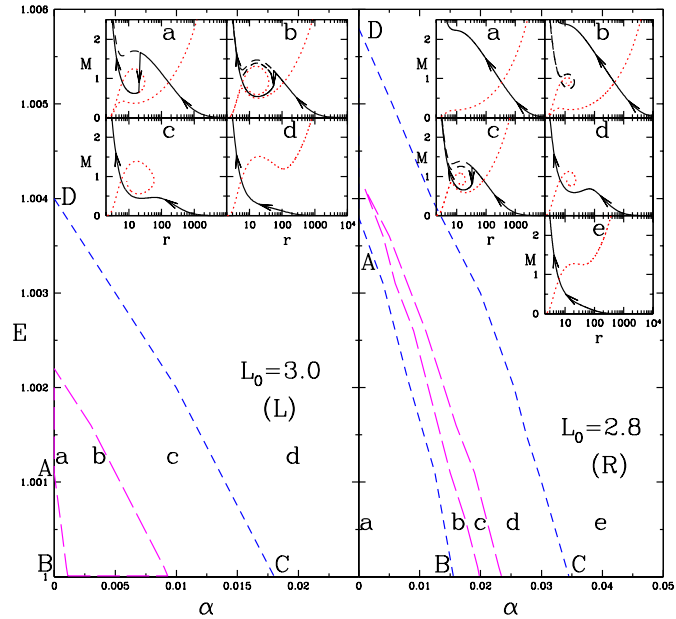}
 \caption{Parameter space of $E~{\rm and}~\alpha$ for given values of $L_0=3$ (L) and $L_0=2.8$ (R). In the inset panels,
solutions, \ie $M$ versus $r$, are plotted, corresponding to the flow parameters ($E~{\rm and}~\alpha$) from various regions marked as
a-e. In both the plots ABCDA is the region for multiple critical points.}
 \label{fig:fig6}
\end{figure}
In Fig. (\ref{fig:fig6} L), we plot the parameter space of $E~{\rm and}~\alpha$ for $L_0=3$ and various regions
in the parameter space are marked as a---d and the typical solutions are plotted in the inset
marked by the same alphabets. In Fig.(\ref{fig:fig6}R),
we plot $E~{\rm and}~\alpha$ for $L_0=2.8$ and various regions are marked as a---e, and the corresponding solutions are plotted
in inset panels. Therefore, parameter space depicted in Figs. (\ref{fig:fig6} L \& R) is analogous to the parameter space
depicted in Fig. (\ref{fig:fig4}), which pans all possible
accretion solutions. It may be noted that the solutions for {$\alpha=0$} which harbour shocks also exhibit steady shocks
up to moderate levels of $\alpha$, but solutions which were Bondi type to start with for $\alpha=0$,
generate a shock transition above a critical value of $\alpha$. For these kind of solutions, one can identify
two critical viscosity parameters, one denotes the onset of steady shock, and the other which marks the limit above
which no steady shock is obtained.

\subsection{Outflow solutions} \label{subsec:outsoln}

\begin{figure}
 \centering
 \includegraphics[width=9.0cm]{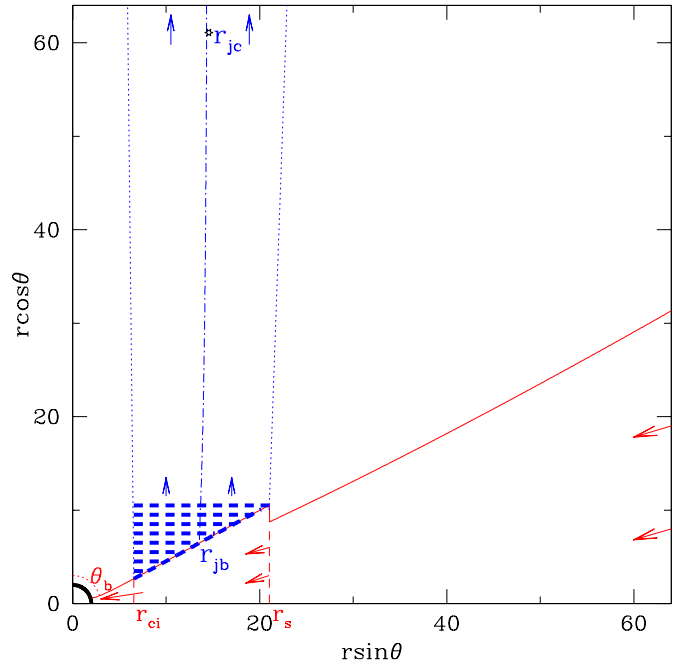}
 \caption{Typical accretion-jet flow geometry for accretion disc parameters $E=1.0001, L_0=2.92, \alpha=0.01$ and $\xi=1$.
Here solid (red) curve represents disc-half {height}. Dot-dashed (blue) line
is jet stream-line for von Zeipel parameter $Z_\phi=13.28$, and dotted {(blue)} line is the inner and outer boundary of jet flow
cross-section.
The jet sonic point is located at $r_{\rm jc}$. Arrows represent direction of bulk motion and the solid thick quarter of a circle
represents the event horizon.}
 \label{fig:fig7}
\end{figure}

It has been shown in many simulations that the PSD drives bipolar outflows \citep{msc96,mrc96,dcnm14}, 
and in theoretical studies of simultaneous accretion-ejection model in the pNp regime,
the flow geometry
of the bipolar outflow or jet was considered within the two surfaces, one, centrifugal barrier surface (pressure maxima)
and the other, funnel wall (minima of the effective potential), both described in the off-equatorial region
\citep{cd07,kscc13,kcm14}. The problem is that both these surfaces depend
primarily on the angular momentum of the flow, and therefore the outflow geometry depends poorly
on the base of the jet or other factors of the flow, which should not be the case. In order to circumvent this as
well as in GR, we were forced by correct physics to obtain the local outflow cross-section by identifying
the relevant VZS, which is not bound by the limitations of pNp regime. In Fig. (\ref{fig:fig7}), we present
the flow geometry of accretion disc, as well as bipolar jets which are actually solved self-consistently for accretion disc
parameters $E=1.0001, L_0=2.92, \alpha=0.01$, where the disc half-height is plotted as solid curve, the jet streamline
is represented by dot-dashed curve, while the dotted curve shows the jet flow geometry. The arrows show the
direction of the flow.

\begin{figure}
 \centering
 \includegraphics[width=9.0cm]{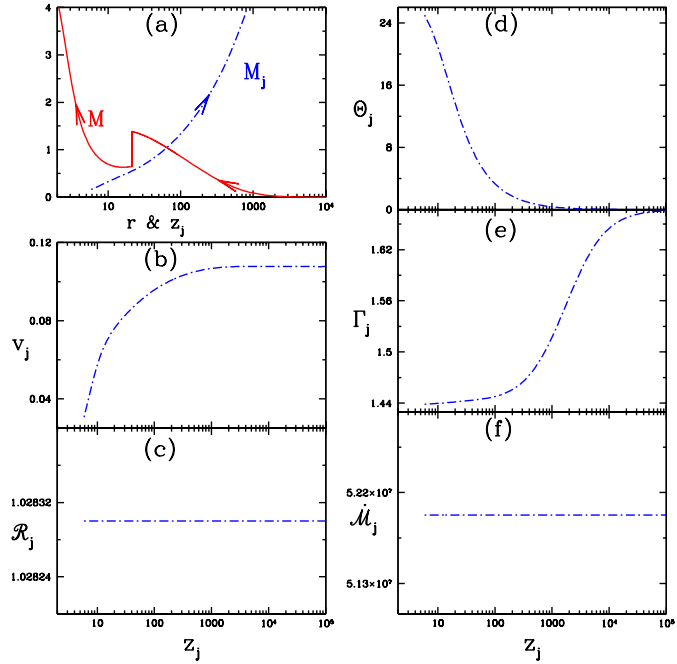}
 \caption{(a) Accretion Mach number $M$ (solid) is plotted w.r .t $r$ and jet Mach number $M_j$ (dashed-dot) is plotted
w. r. t $z_j$; (b) variation jet 3-velocity $v_j$; (c) jet Bernoulli parameter ${\Re}_j$;
(d) jet dimensionless temperature $\Theta_j$; 
 (e) jet adiabatic index $\Gamma_j$ and (f) jet entropy ($\dot{\cal M}_j$) all are plotted w.r.t $z_j$.
Accretion disc parameters are 
 $E=1.001,~ L_0=2.906,~ \alpha=0.01$. The disc and jet flow composition is described by
$\xi= 1.0$ and relative mass outflow rate is $R_{\dot{m}}=0.053$.}
 \label{fig:fig8}
\end{figure}

In Fig. (\ref{fig:fig8}a), we plot the combined accretion-jet solution, here the accretion Mach number $M$ (solid)
is plotted with respect to $r$, while the jet Mach number $M_{\rm j}$ is plotted w.r.t $z_{\rm j}$ in the same panel. In Figs.
(\ref{fig:fig8}b-f), we plot various jet variables, for \eg~ the jet three-velocity $v_{\rm j}$ (Fig. \ref{fig:fig8}b),
$\Re_{\rm j}$ (Fig. \ref{fig:fig8}c), $\Theta_{\rm j}$ (Fig. \ref{fig:fig8}d), $\Gamma_{\rm j}$ (Fig. \ref{fig:fig8}e),
and $\dot{\cal M}_{\rm j}$ (Fig. \ref{fig:fig8}f), for accretion disc parameters $E=1.001,~ L_0=2.906\mbox{ and } \alpha=0.01$.
The jet is followed up to $z_{\rm j}=10^4\rg$ above the equatorial plane of the accretion disc. In Schwarzschild metric we do not
find multiple sonic points in jets, and jets are transonic flow also with only one sonic point. However, the jet achieves
fairly high terminal speed ($\sim 0.11c$), inspite of being only thermally driven
(\ie~ $v_{\rm j}$ increases as $\Theta_{\rm j}$ decreases). The specific energy of the jet $\Re_{\rm j}$
and its entropy-accretion rate
${\dot {\cal M}}_{\rm j}$ are constants of motion since the jet is assumed to be adiabatic.

\begin{figure}
 \centering
 \includegraphics[width=9.0cm]{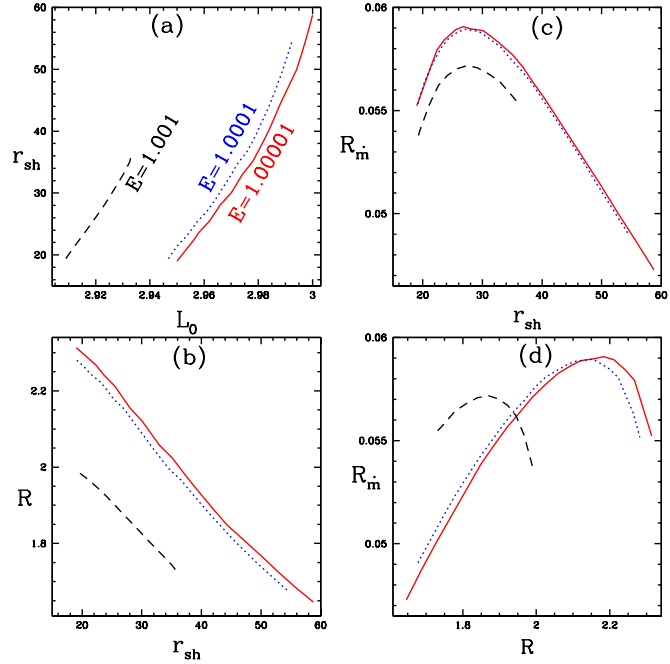}
 \caption{(a) Variation of shock location $\rsh$ with $L_0$;(b) compression
ratio $R$ with $\rsh$; (c) mass outflow rate $\rmdot$ with $\rsh$ and (d) $\rmdot$ with $R$. Each curve is for $E=1.00001$
(solid, red),
$1.0001$ (dotted, blue) and $1.001$ (dashed, black).
For all the cases, $\xi=1.0$ and $\alpha=0.01$.}
 \label{fig:fig9}
\end{figure}

We have shown in Figs. \ref{fig:fig2}---\ref{fig:fig4} that for given values of $\alpha$, the nature of accretion solution
depends on $L_0$ and since accretion disc launches the jet, we would like to analyse how the jet depends on the
inner boundary condition of the flow. In Fig. (\ref{fig:fig9}a), we plot $\rsh$ as a function of $L_0$; each curve
is obtained for a given value of $E=1.00001$ (solid, red), $E=1.0001$ (dotted, blue) and $1.001$ (dashed, black).
The viscosity is given by $\alpha=0.01$ and the disc-jet is composed of electron-proton fluid. For a given value of $L_0$,
the $\rsh$ increases with increasing $E$ if steady shock is allowed by the flow, while for a given value
of $E$, $\rsh$ increases with $L_0$. The corresponding compression ratio $R$
as a function of $\rsh$ is shown in Fig. (\ref{fig:fig9}b), while the relative mass outflow rates $\rmdot$ (\eg~ equation \ref{rmd1.eq})
are plotted with $\rsh$ in
Fig. (\ref{fig:fig9}c). As the $\rsh$ increases, the compression ratio decreases (Fig. \ref{fig:fig9}b)
so the upward thrust becomes weaker. However, higher value of $\rsh$ also makes the surface area of PSD and therefore the base of the
jet larger, so the net mass flowing out as jet should become more. These contradictory tendencies cause the mass outflow rate
to peak at some intermediate value of $\rsh$, as well as that of $R$ (Fig. \ref{fig:fig9}d).
\begin{figure}
 \centering
 \includegraphics[width=9.0cm]{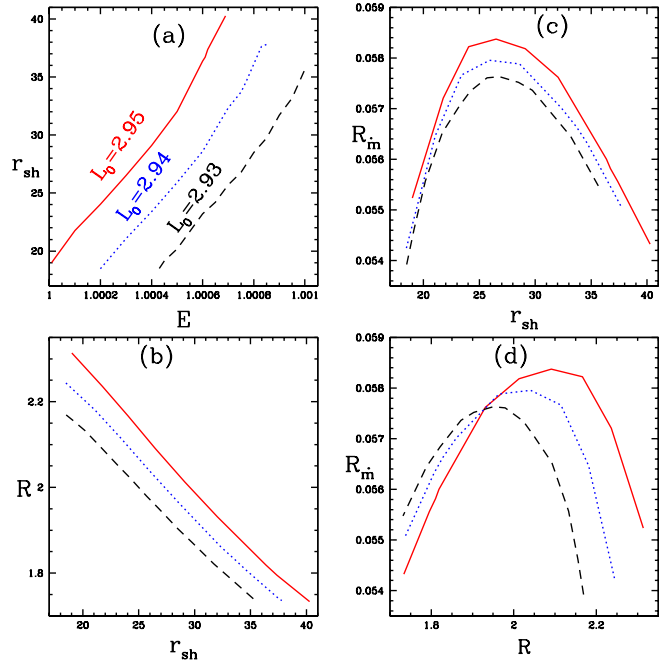}
 \caption{(a) Variation of $\rsh$ with $E$, (b) $R$ with $\rsh$, (c) $\rmdot$ with $\rsh$
and (d) $\rmdot$ with $R$. Each curve is plotted 
for $L_0=2.95$ (solid, red), $2.94$ (dotted, blue) and 
 $2.93$ (dashed, black). For all the curves, $\xi=1.0~{\rm and}~ \alpha=0.01$.}
 \label{fig:fig10}
\end{figure}

In Figs. (\ref{fig:fig10}a-d), the converse dependence is studied where, $\rsh$ is plotted with $E$, where
each curve represent $L_0=2.95$ (solid, red), $2.94$ (dotted, blue) and 
$2.93$ (dashed, black). The composition of the flow and the viscosity parameter is the same as in Fig. (\ref{fig:fig9}a-d).
The shock location increases (Fig. \ref{fig:fig10}a) with both $L_0$ and $E$, as was observed in the previous figure.
As the shock increases, the compression ratio decreases (Fig. \ref{fig:fig10}b). However, $\rmdot$ do not monotonically
increase with decreasing $r_s$, for the same reason as was discussed in the previous figure.
Interestingly, lower $L_0$ produces lower values of $\rsh$, but since these shocks are mainly rotation mediated,
so lower $L_0$ implies weaker shock, and therefore the compression ratio $R$ ($\equiv$ the amount of squeezing on the
post-shock flow) is weak too. Therefore, although the shock is located closer
to the BH for lower $L_0$, the $\rmdot$ is less even for the same values of $\rsh$. 

\subsubsection{Effect of viscosity, $\alpha$}\label{subsubsec:eov}
\begin{figure}
 \centering
 \includegraphics[width=9.0cm]{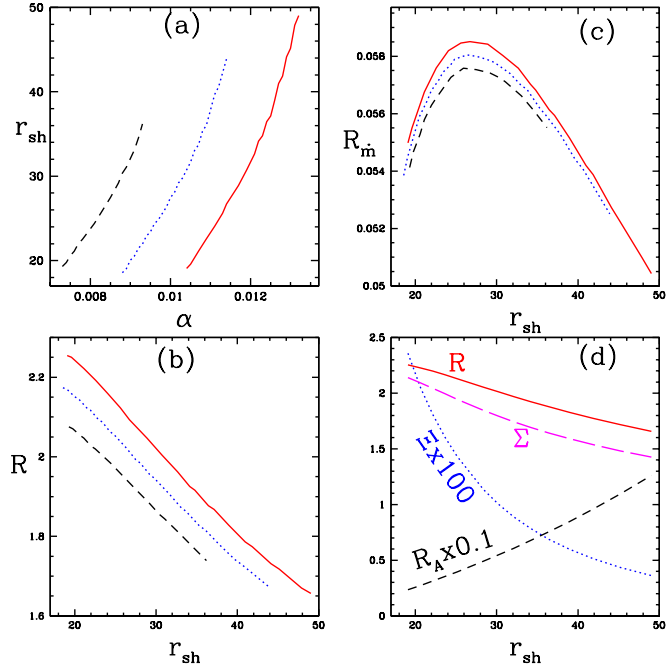}
 \caption{Variation of $\rsh$ with $\alpha$ (a), $R$
 with $\rsh$ (b), $\rmdot$ with $\rsh$ (c). 
Each curve is for $E=1.0001$ (solid, red), $E=1.00055$ (dotted, blue) and $E=1.001$ (dashed, black).
In panel (d), we plot $R$ (solid, red), $\Sigma$ (long dashed, magenta), $\varXi$ (dotted, blue)
$R_{\rm A}$ (dashed, black) for $E=1.0001$ (solid, red curve of panels a---c).
For all curves $\xi=1.0, ~L_0=2.94$. Dependence of $\rsh$ on $\alpha$, for fixed inner boundary condition.}
 \label{fig:fig11}
\end{figure}

In Fig. (\ref{fig:fig11}a), we plot how $\rsh$ would behave with the change in $\alpha$, for fixed inner boundary condition
or for the same values of $E$ and $L_0$. We plot the corresponding
$R$ as a function of $\rsh$ (Fig. \ref{fig:fig11}b) and $\rmdot$ with $\rsh$ (Fig. \ref{fig:fig11}c).
Each curve is for constant $E=1.0001$ (solid, red), $E=1.00055$ 
(dotted, blue) and $E=1.001$ (dashed, black), where for all curves $\xi=1.0, ~L_0=2.94$.
And in Fig. (\ref{fig:fig11}d), we plot $R$ (solid, red), $\Sigma$ (long dashed, magenta), $\varXi$ (dotted, blue)
and $R_{\rm A}$ (dashed, black) for $E=1.0001$ (solid, red curve of Fig. \ref{fig:fig11}a---c).
Since $E$ is a constant of motion for the accretion disc, and $L_0$ is the bulk angular momentum on the BH
horizon, so fixed values of $E$ and $L_0$ correspond to fixed inner boundary condition. For same $L_0$ and $E$,
as one increases $\alpha$, then the angular momentum at the outer edge of the disc would be higher. This implies that in the
PSD too, the angular momentum $L$ or specific angular momentum $\lambda$ will be higher. Thus the shock location would
increase with $\alpha$. For a given $E$, the compression ratio decreases with increasing $\rsh$. 
Since the accretion shock is rotation dominated,
therefore, the $\rsh$ will increase for hotter flow ($\equiv$ higher $E$), but the compression ratio will decrease.
Thus, for a given value of $\alpha$, $\rmdot$ will be less for higher values of $E$.
Fig. (\ref{fig:fig11}c) shows that the $\rmdot$ is low for high and low values of $\rsh$ and maximizes at some
intermediate value. In Fig. (\ref{fig:fig11}d), we find out why the mass outflow rate or $\rmdot$ has a non-uniform
dependence on $\rsh$. From equation (\ref{rmd.eq}), we know that $\rmdot$ increases with increasing $R_{\rm A}$, $R$ and $\varXi$,
but decreases with increasing $\Sigma$. So as the $\rsh$ increases (Fig. \ref{fig:fig11}a), Fig. (\ref{fig:fig11}d)
shows that $R$ and $\varXi$ decrease, which implies that the post-shock
thrust which is responsible for driving the jet decreases which should decrease $\rmdot$.
However, $R_{\rm A}$, or the ratio between jet cross-sectional area and the PSD surface area,
increases; therefore, this should increase $\rmdot$.
These two contradictory tendencies, make $\rmdot$ attain low values when the $\rsh$ is very close to horizon
and when it is far away, but maximize for some intermediate values.  
\begin{figure}
 \centering
 \includegraphics[width=9.0cm]{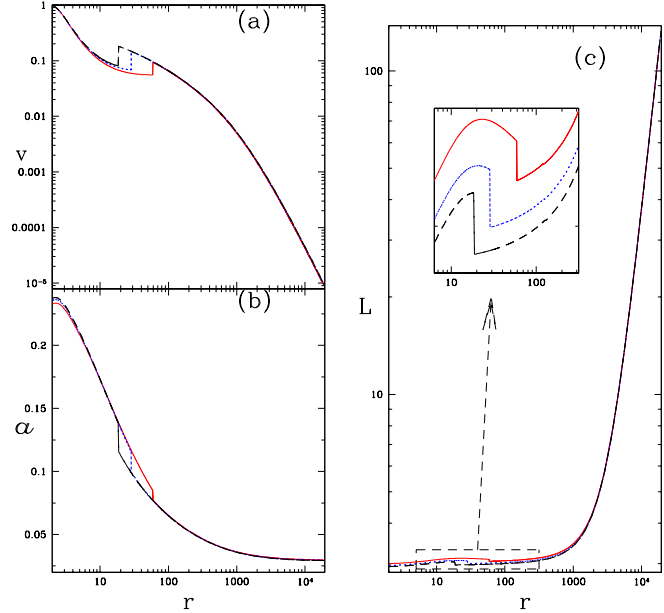}
 \caption{Three-velocity $v$ (a), local sound speed {$a$} (b) and bulk angular momentum $L$ (c)
of the accretion disc plotted with $r$. Each curve is for $\alpha=0.01$ (solid, red), $\alpha=0.0105$ (dotted, blue)
and $\alpha=0.011$ (dashed, black). For all the curves,
the outer boundary is at $\rout=19835.3\rg$, the corresponding  
specific angular-momentum is the Keplerian angular momentum at $\rout$, \ie $\lambda_{\rm out}=\lambda_{\rm K}=140.85$
and the constant of motion for all the curves is $E=1.0001$. Inset in panel (c) zooms on the $L$ distribution around the
location of the shock.}
 \label{fig:fig12}
\end{figure}

Let us compare the flow variables of accreting matter which starts with the same outer boundary condition.
We plot and compare the three velocity $v$ (Fig. \ref{fig:fig12}a), sound speed $a$ (Fig. \ref{fig:fig12}b)
and the bulk angular momentum $L$ (Fig. \ref{fig:fig12}c) of accretion flows starting
with the same outer boundary condition $E=1.0001$
and $\lambda_{\rm out}=\lambda_{\rm K}=140.85$ at the outer edge of the accretion disc $\rout=19835.3$. Each curve represents
the solution for $\alpha=0.01$ (solid, red), $\alpha=0.0105$ (dotted, blue)
and $\alpha=0.011$ (dashed, black), and the net relative mass outflow computed were $\rmdot=0.047$ (solid, red),
$\rmdot=0.059$ (dotted, blue) and $\rmdot=0.054$ (dashed, black). As $\alpha$ is increased, the net angular momentum
of the inner disc decreases, and since the shock is rotation driven, lower angular momentum causes $\rsh$
to decrease (see the inset of Fig. \ref{fig:fig12}c).
\begin{figure}
 \centering
 \includegraphics[width=9.0cm]{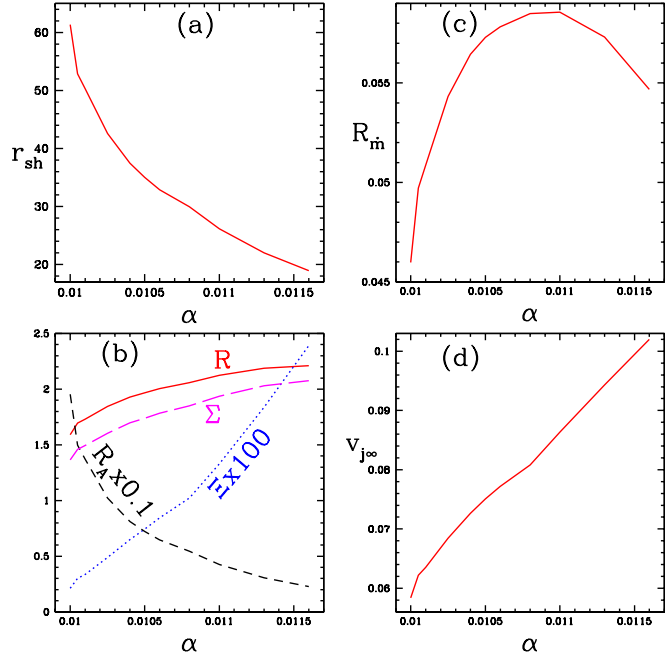}
 \caption{(a) Variation of $\rsh$ with $\alpha$, (b) $R$ (solid, red), $\varXi$ (dotted, blue), $\Sigma$ (long dashed,
 magenta) and $R_{\rm A}$
(dashed, black) with $\alpha$, (c) $\rmdot$
 with $\alpha$ and (d) $v_{{\rm j}\infty}$
with $\alpha$.
The outer boundary is at $\rout=16809.016$, and corresponding  
specific angular momentum is the Keplerian angular momentum at $\rout$, \ie~ $\lambda_{\rm out}=\lambda_{\rm K}=129.662$. 
For all the curves, $E=1.0001,~ \xi=1.0$. Dependence of $\rsh$ on $\alpha$ for fixed outer boundary.}
 \label{fig:fig13}
\end{figure}

Although it is interesting to show how $\alpha$ will affect $\rsh$, for the same inner boundary condition
of the disc. But
the physics of accretion disc is controlled by outer boundary condition, so it will be more physical to study how the disc
solution, as well as, the ensuing jet solutions depend on $\alpha$ when the outer boundary condition of the
accretion disc is kept the same. In Fig. (\ref{fig:fig13}a), $\rsh$ is plotted with $\alpha$ for $E=1.0001$. The outer
boundary of the disc is $\rout=16809.016$ for all solutions for which the curve is plotted. The specific angular momentum
at $\rout$ is the local Keplerian value $\lambda_{\rm out}=\lambda_{\rm K}=129.662$. Since $E$ is a constant of motion
for all the solutions presented, and $\lambda_{\rm out}$ is also same for all the disc solutions, so comparing
solutions for same $E$ and $\lambda_{\rm out}$ is equivalent to comparing solutions starting with the same outer boundary.
Viscosity transports angular momentum outwards;
therefore, for a given value of $E$, the shock moves closer to the BH as viscosity is increased.
So $\rsh$ decreases with increasing $\alpha$. 
The corresponding dependence of $R$ (solid, red), $\varXi$ (dotted, blue), $\Sigma$ (long dashed, magenta) and $R_{\rm A}$
(dashed, black) with $\alpha$ has been plotted in Fig. \ref{fig:fig13}b.
The shock becomes stronger as it moves towards the horizon therefore $R$ increases, but the enhanced
compression also squeezes more matter along the jet channel so $\varXi$ increases too.
However, $\Sigma$ increases and $R_{\rm A}$ decreases which should decrease the $\rmdot$. Such antagonistic tendencies
make the $\rmdot$ to peak at some intermediate $\alpha$, as is depicted in Fig. (\ref{fig:fig13}c).
In Fig. \ref{fig:fig13}d, the jet terminal speed $v_{{\rm j}\infty}$
with $\alpha$ is plotted. Since $R$ increases, so the upward thrust also increases, making jets
stronger, even if $\rmdot$ decrease.
It means we can have stronger but lighter jets.
\subsubsection{Effect of composition, $\xi$} \label{subsubsec:eoc}
\begin{figure}
 \centering
 \includegraphics[width=9.0cm]{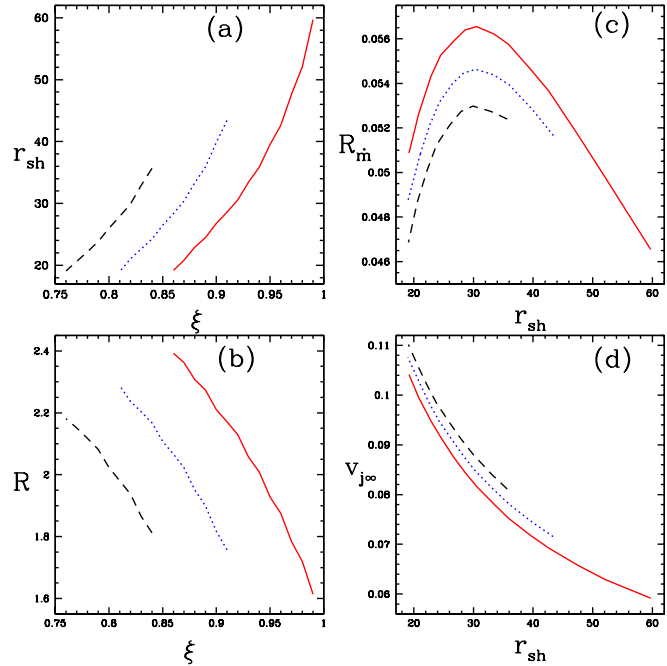}
 \caption{Dependence of $\rsh$ on composition parameter $\xi$ (a), $R$ with $\xi$ (b), $\rmdot$
 on $\rsh$ (c) and $v_{{\rm j}\infty}$ with $\rsh$ (d). Each plot corresponds 
 to $E=1.0001$ (solid, red), $1.00055$ (dotted, blue) and $1.001$ 
 (dashed, black).
 For all the curves $L_0=3.0,~ \alpha=0.01$.}
 \label{fig:fig14}
\end{figure}

In all the previous figures, we dealt with fluid composed of only electrons and protons. \citet{cr09}
showed that if the proton proportion is reduced (where the charge balance is maintained by proportionate increase
of positrons), the flow becomes thermally more relativistic because the decrease in thermal energy is compensated by 
decrease in inertia of the flow. Fig. (\ref{fig:fig14}a) shows that $\rsh$ increases with $\xi$, where each curve is for
$E=1.0001$ (solid, red), $1.00055$ (dotted, blue) and $1.001$ 
 (dashed, black), and $L_0=3.0$ and $\alpha=0.01$. Higher $\rsh$ implies lower $R$ (Fig. \ref{fig:fig14}b);
as a result, $\rmdot$ decrease with increasing $\rsh$, although, due to the related increase in the jet base and other factors
[dealt with related to Figs. \ref{fig:fig11}(a)-(d)], $\rmdot$ peaks
at some intermediate value (Fig. \ref{fig:fig14}c). In Fig. (\ref{fig:fig14}d), the terminal speed of the jet $v_{{\rm j}\infty}$
with $\rsh$ is plotted. As the shock recedes, the speed of the jet decreases, even where $\rmdot$ is increasing.
But if the accretion disc flow is more energetic, the jet terminal speed is higher, although $\rmdot$ is lower.

\begin{figure}
 \centering
 \includegraphics[width=9.0cm]{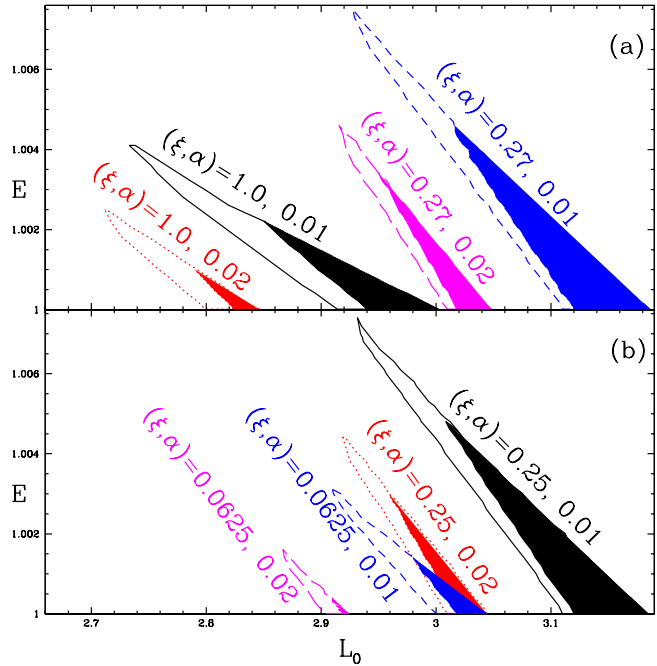}
 \caption{(a) Comparison of the shock parameter space in $E-L_0$ with mass-loss (shaded area) and without mass-loss
(bounded area) for different disc parameters, $\xi, \alpha=1.0, 0.01$ (solid), $1.0, 0.02$ (dotted), $0.27, 0.01$ (dashed)
and $0.27, 0.02$ (long dashed). (b) Comparison of shock parameter space $E-L_0$ for
$(\xi, \alpha)=0.25, 0.01$ (solid), $0.25, 0.02$ (dotted) and $0.0625, 0.01$ (dashed). Same notations for shocked region
with or without mass-loss.}
 \label{fig:fig15}
\end{figure}

In Figs. \ref{fig:fig15}(a) and b), we plot the shock parameter space in the $E-L_0$ space for various combinations
of viscosity and composition parameters like $\xi, \alpha=1.0, 0.01$ (solid), $1.0, 0.02$ (dotted), $0.27, 0.01$ (dashed)
and $0.27, 0.02$ (long dashed) in Fig. \ref{fig:fig15}(a)
and for $(\xi, \alpha)=0.25, 0.01$ (solid), $0.25, 0.02$ (dotted) and $0.0625, 0.01$ (dashed) in Fig. \ref{fig:fig15}(b).
The shaded region indicates the steady shock region of the parameter space when mass-loss is considered.
Similar to the inviscid study \citep{cc11}, the shock parameter space moves to the higher energy
direction of the parameter space till $\xi$ is reduced from $1$ to $0.27$. As $\xi$ is reduced further,
the shock parameter space moves towards the low-energy side. The reduction of steady shock parameter space due to mass-loss
actually indicates that shock in accretion actually exists in a wide range, but only as a time-dependent one.

\section{Discussions and Conclusions} \label{sec:conclusn}
Investigations of viscous accretion disc around a BH in general relativistic regime are important,
because a BH is necessarily
a relativistic object, and close to the horizon the departure from Newtonian description is significant.
Analysis with Paczy\'nskii-Wiita pNp gives us an overall qualitative understanding, but 
quantitatively the results or predictions based on pNp are bound to be wrong. One of the artefacts of pNp is that the
disc closer to the horizon is hotter than they actually are. Moreover, matter speed exceeds the speed of light.
Furthermore, the relativistic shear tensor is more complicated than the Newtonian variant. All these reasons
influenced us to re-investigate the self-consistent, simultaneous, viscous
accretion-ejection solutions around BHs in the general relativistic regime. We chose the simplest
type of BH for our study, simply because the influence of Kerr metric in powering jets
is not very conclusive in some of the
studies made on microquasars \citep{f10}.

In this paper, we first analysed only the accretion solution (\ie~assuming $\rmdot=0$).
The solution procedure was influenced by the methodology laid down by Becker and his collaborators in the pNp regime
in order to find the sonic points self-consistently.
We have generalized it for general relativistic domain. Moreover, the equation of the state of flow
is also relativistic and not a Newtonian polytropic gas, as was used by Becker. 
We obtained all possible
accretion solutions in the advective domain, starting from Bondi type (single $\rco$-type sonic point), solutions
where multiple sonic points formed, shocked solution and ADAF solutions (single $\rci$-type sonic points).
All of these various solutions were obtained for various combinations of $E,~L_0~{\rm and}~\alpha$, so in this respect each solution
is a three-parameter class of solutions. A quick comparison shows that, in the pNp regime shocks are stable for
higher viscosity parameters
($\alpha \sim 0.3$),
but in the general relativistic regime (solutions of this paper) shock remains steady for much lower values
($\alpha \lsim 0.06$).
In the pNp regime, the shear tensor is much simpler, and the $r-\phi$ component is proportional to the radial derivative
of the angular velocity \citep{kc13}, but in the GR regime the shear tensor is much more complicated and
depends on four-velocities and
their derivatives, as well as various components of 
four-acceleration. The approximated version used in this paper also has radial four-velocity term in addition to the
derivative of azimuthal component of four-velocity (see equation \ref{rphi.eq}). As a result, the shock is made unstable for lower values of
$\alpha$. Instability of shock is actually good because not only such oscillation explains quasi-preiodic oscillaitons (QPOs)
but also does
additional pumping to
generate stronger jet \citep{dcnm14}.

In this paper, we did a detailed study of jet morphology based on the works of \citet[and citations therein]{c85}.
Here too, physics in curved space-time differed from Newtonian or pNp regime.
In GR, the entropy constant surface in non-equatorial plane for adiabatic flow
coincides with constant angular momentum surfaces called VZS which are not cylinders
of flat space-time. \citet{c85} showed
ways to relate these surfaces with angular momentum of the flow. Since jets are tenuous and are likely to be adiabatic
till they interact with the ambient medium, so adiabatic jet is a fairly good assumption.
Therefore, VZS becomes the natural streamline of the flow.
Interestingly, the accreting matter has predominantly the radial and azimuthal component of four-velocity, and can be
quite accurately described about the equatorial plane; however, the jet has all three components of the four-velocity.
We turned this essentially three spatial dimensional problem into an effective one-dimensional problem, by projecting
jet equations of motion along the streamline defined by VZS and the methodology is properly defined in Section
\ref{subsec:ouequatn} and Appendix \ref{app:gpp}.
The streamline obtained in such a manner is very rich, where some combination of parameters produced partially
pinched-off streamlines, which are ripe case for multiple sonic points and internal shocks in jets.
However, in Schwarzschild  geometry
above the equatorial plane, such jet geometry cannot be connected to the inner disc region of the accreting matter.
Only those VZS solutions which did not allow multiple sonic points were found to be relevant, and therefore
we only obtained monotonic jet solutions. So depending on where the shock in accretion disc forms, the jet base and
jet flow geometry change. This fact is also markedly different from pNp prescription, since in pNp prescription
the jet geometry is weakly dependent on the jet base or other accretion disc properties.

The response of the shock location with viscosity is similar to our studies in pNp. Even
relativistic viscosity makes the accretion shock to move closer to the event horizon as it is increased,
if the outer boundary condition of the accretion disc is kept the same. As the shock moves closer to the BH,
compression ratio across the shock increases, which generally increases the jet strength. This initially increases
the relative mass outflow rate, but the mass outflow rate starts to drop, as the shock moves closer to the BH.
We also showed that as the shock becomes stronger (forms closer to the BH), it also increases the
upward thrust (Figs. \ref{fig:fig11}d, \ref{fig:fig13}b).
Independent of how much percentage of accreting matter is pumped out as jets, because of the increased thrust
the terminal speed of the
jet material increases with decreasing shock location.
The shock cannot move too close
to the BH and still remain stationary, at some point it will become unstable and start to oscillate. And that
is marked by the closest limit up to which we found steady shocks in this analysis. Although
this investigation is steady state and cannot conclusively comment on essentially time-dependent phenomena,
but it can be conjectured that increasing
viscosity even more should increase the oscillation frequency of the shock too.
Such a situation does mimic an outbursting source, where it had been shown that as the object moves from low hard
state to intermediate states, the QPO as well as the jet strength increases.
We also showed that fluid described by a proton proportion of $27\%$ of the electron number density (the rest being
positrons), or $\xi=0.27$ is thermally the most relativistic, and can form shocks in accretion in the largest portion
of the $E-L_0$ parameter space. We also showed that shock in accretion does not form
if the accretion disc is composed of pair plasma.
And the mass outflow rate is less than $6\%$ of the accretion rate for any value of $\xi$.
We also showed that independent of the composition of the flow, hotter the accretion disc, faster will be the
emanating jet, although the mass outflow rate may not be very high.

\section*{Acknowledgements}
The authors acknowledge the anonymous referee for fruitful suggestions to improve the quality of the paper.

{}

\appendix
\section{Calculation of \hp}\label{app:gpp}
The equation of tangent is defined on jet streamline at any point, 
\begin{equation}
 x_p=mr_{\rm j}{\rm sin}\theta_{\rm j}+c_i,
 \label{tngt.eq}
\end{equation}
where $m=\frac{{\rm d}y}{{\rm d}x}=(r_{\rm j}-2){\rm cot}\theta_{\rm j}+(r_{\rm j}-3){\rm tan}\theta_{\rm j}$ and $c_i$
are the slope and intercept of the tangent, respectively. 
The basis vector along jet streamline is defined as,
\begin{equation}
 {\bf e}_p=(\frac{\partial r_j}{\partial x_p})){\bf e}_r+(\frac{\partial \theta_j}{\partial x_p}){\bf e}_\theta,
 \label{bsvec.eq}
\end{equation}
where, ${\bf e}_p=h_p\hat{e}_p, {\bf e}_r=h_r\hat{e}_r$ and ${\bf e}_\theta=h_\theta\hat{e}_\theta$. Here, $\hat{e}_p, \hat{e}_r$ and 
$\hat{e}_\theta$ are unit basis vectors along tangent, radial and polar direction, respectively. 
The magnitude of this basis vector is defined as,
\begin{equation}
 h_p^2=h_r^2\left(\frac{\partial r_{\rm j}}{\partial x_p}\right)^2+h_\theta^2\left(\frac{\partial\theta_{\rm j}}{\partial x_p}\right)^2,
 \label{hp1.eq}
\end{equation}
where $h_r^2=g_{rr}=(1-2/r_{\rm j})^{-1}$ and $h_\theta^2=g_{\theta\theta}=r_{\rm j}^2$ are metric components. 
In order to obtain $h_p$, 
we have to take partial differentiation with respect to $x_p$ of equations (\ref{VZp.eq}) and (\ref{tngt.eq}), we get
\begin{eqnarray}\nonumber
 \left[\frac{(r_{\rm j}-3)}{r_{\rm j}(r_{\rm j}-2)}\right]^2\left(\frac{\partial r_{\rm j}}{\partial x_p}\right)^2=
 {\rm cot}^2\theta_{\rm j}\left(\frac{\partial\theta_{\rm j}}
 {\partial x_p}\right)^2 ~~~~~~~\mbox{and}~~\\ {\rm cos}^2\theta_{\rm j}=[2r_{\rm j}-2-{\rm sin}^2\theta_{\rm j}]^2\left(\frac{\partial r_{\rm j}}
 {\partial x_p}\right)^2+
 r_{\rm j}^2{\rm tan}^2\theta_{\rm j}[r_{\rm j}-4+{\rm sin}^2\theta_{\rm j}]^2\left(\frac{\partial\theta_{\rm j}}
 {\partial x_p}\right)^2.
 \label{bsisdv.eq}
\end{eqnarray}
Now, expression of $h_p$ is obtained by using equation (\ref{bsisdv.eq}) in equation (\ref{hp1.eq}), 
\begin{equation}
h_p^2=\left(1-\frac{2}{r_{\rm j}}\right)^{-1}\left[\frac{{\rm cos}^2\theta_{\rm j}
+{{\rm sin}^2\theta_{\rm j}(r_{\rm j}-3)^2}/({r_{\rm j}(r_{\rm j}-2)})}{\{2r_{\rm j}-2-
 {\rm sin}^2\theta_{\rm j}\}^2+\{{\rm tan}^2\theta_{\rm j}(r_{\rm j}-4+{\rm sin}^2\theta_{\rm j})({r_{\rm j}-3})/({r_{\rm j}-2})\}^2}\right]
 \label{hp.eq}
\end{equation}

\end{document}